# Title: Transcription factor clusters regulate genes in eukaryotic cells

**Authors:**

Adam J. M. Wollman[a,1], Sviatlana Shashkova[a,b,1], Erik G. Hedlund[a], Rosmarie Friemann[b], Stefan Hohmann[b,c], Mark C. Leake[a,2]

**Affiliations:**

[a] Biological Physical Sciences Institute, University of York, York YO10 5DD, UK

[b] Department of Chemistry and Molecular Biology, University of Gothenburg, 40530 Göteborg, Sweden.

[c] Department of Biology and Biological Engineering, Chalmers University of Technology, 41296 Göteborg, Sweden.

[1] A.J.M.W. and S.S. contributed equally to this work.

[2] To whom correspondence should be addressed. Email: mark.leake@york.ac.uk.

Corresponding author: Prof Mark Leake, Biological Physical Sciences Institute, University of York, York YO10 5DD, UK. Tel: +44 (0)1904322697. Email: mark.leake@york.ac.uk. Orcid ID**:** http://orcid.org/0000-0002-1715-1249.

**Keywords:** gene expression; transcription factors; single-molecule; super-resolution; cell signaling.

**Abstract**
Transcription is regulated through binding factors to gene promoters to activate or repress expression, however, the mechanisms by which factors find targets remain unclear. Using single-molecule fluorescence microscopy, we determined *in vivo* stoichiometry and spatiotemporal dynamics of a GFP tagged repressor, Mig1, from a paradigm signaling pathway of *Saccharomyces cerevisiae*. We find the repressor operates in clusters, which upon extracellular signal detection, translocate from the cytoplasm, bind to nuclear targets and turnover. Simulations of Mig1 configuration within a 3D yeast genome model combined with a promoter-specific, fluorescent translation reporter confirmed clusters are the functional unit of gene regulation. *In vitro* and structural analysis on reconstituted Mig1 suggests that clusters are stabilized by depletion forces between intrinsically disordered sequences. We observed similar clusters of a co-regulatory activator from a different pathway, supporting a generalized cluster model for transcription factors that reduces promoter search times through intersegment transfer while stabilizing gene expression.

**Introduction**

Cells respond to their environment through gene regulation involving protein transcription factors. These proteins bind to DNA targets of a few tens of base pairs (bp) length inside ~500-1,000bp promoter sequences to repress/activate expression, involving single (1) and multiple (2) factors, resulting in the regulation of target genes. The mechanism for finding targets in a genome ~six orders of magnitude larger is unclear since free diffusion followed by capture is too slow to account for observed search times (3). Target finding may involve heterogeneous mobility including nucleoplasmic diffusion, sliding and hops along DNA up to ~150bp, and even longer jumps separated by hundreds of bp called intersegment transfer (4–6).

In eukaryotes, factor localization is dynamic between nucleus and cytoplasm (7). Although target binding sites in some cases are known to cluster in hotspots (8) the assumption has been that factors themselves do not function in clusters but as single molecules. Realistic simulations of diffusion and binding in the complex milieu of nuclei suggest a role for multivalent factors to facilitate intersegment transfer by enabling DNA segments to be connected by a single factor (9).

The use of single-molecule fluorescence microscopy to monitor factor localization in live cells has resulted in functional insight into gene regulation (10). Fluorescent protein reporters, in particular, have revealed complexities in mobility and kinetics in bacterial (11) and mammalian cells (12) suggesting a revised view of target finding (4).

Key features of gene regulation in eukaryotes are exemplified by glucose sensing in budding yeast, *Saccharomyces cerevisiae*. Here, regulation is achieved by factors which include the Mig1 repressor, a Zn finger DNA binding protein (13) that acts on targets including *GAL* genes (14). Mig1 is known to localize to the nucleus in response to increasing extracellular glucose (15), correlated to its dephosphorylation (16). Glucose sensing is particularly valuable for probing gene regulation since the activation status of factors such as Mig1 can be controlled reproducibly by varying extracellular glucose. Genetic manipulation of the regulatory machinery is also tractable, enabling native gene labeling with fluorescent reporters for functioning imaging studies.

We sought to explore functional spatiotemporal dynamics and kinetics of gene regulation in live *S. cerevisiae* cells using its glucose sensing pathway as a model for signal transduction. We used single-molecule fluorescence microscopy to track functional transcription factors with millisecond sampling to match the mobility of individual molecules. We were able to quantify composition and dynamics of Mig1 under physiological and perturbed conditions which affected its possible phosphorylation state. Similarly, we performed experiments on a protein called Msn2, which functions as an activator for some of Mig1 target genes (17) but controlled by a different pathway. By modifying the microscope we were also able to determine turnover kinetics of transcription factors at their nuclear targets.

The results, coupled to models we developed using chromosome structure analysis, indicated unexpectedly that the functional component which binds to promoter targets operates as a cluster of transcription factor molecules with stoichiometries of ~6-9 molecules. We speculated that these functional clusters in live cells were stabilized through interactions of intrinsically disordered sequences facilitated through cellular depletion forces. We were able to mimic those depletion forces in *in vitro* single-molecule and circular dichroism experiments using a molecular crowding agent. Our novel discovery of factor clustering has a clear functional role in facilitating factors finding their binding sites through intersegment transfer, as borne out by simulations of multivalent factors (9); this addresses a long-standing question of how transcription factors efficiently find their targets. This clustering also functions to reduce off rates from targets compared to simpler monomer binding. This effect improves robustness against false positive detection of extracellular chemical signals, similar to observations for the monomeric but multivalent bacterial LacI repressor (4). Our findings potentially reveal an alternative eukaryotic cell strategy for gene regulation but using an entirely different structural mechanism.

**Results**

**Single-molecule imaging reveals *in vivo* clusters of functional Mig1**

To explore the mechanisms of transcription factor targeting we used millisecond Slimfield single-molecule fluorescence imaging (18–20) on live *S. cerevisiae* cells (Fig. 1A and S1). We prepared a genomically encoded green fluorescent protein (GFP) reporter for Mig1 (Table S1). To enable nucleus and cell body identification we employed mCherry on the RNA binding nuclear protein Nrd1. We measured cell doubling times and expression to be the same within experimental error as the parental strain containing no fluorescent protein (Fig. S2A). We optimized Slimfield for single-molecule detection sensitivity with an *in vitro* imaging assay of surface-immobilized purified GFP (21) indicating a brightness for single GFP molecules of ~5,000 counts on our camera detector (Fig. S2B). To determine any fluorescent protein maturation effects we performed cell photobleaching while expression of any additional fluorescent protein was suppressed by antibiotics, and measured subsequent recovery of cellular fluorescence <15% for fluorescent protein components, corrected for any native autofluorescence, over the timescale of imaging experiments (Fig. S2C and S2D).

Under depleted (0-0.2%)/elevated (4%) extracellular glucose (-/+), we measured cytoplasmic and nuclear Mig1 localization bias respectively, as reported previously (15), visible in individual cells by rapid microfluidic exchange of extracellular fluid. (Fig. 1B; SI Appendix). However, our ultrasensitive imaging resolved two novel components under both conditions consistent with a diffuse monomer pool and distinct multimeric foci which could be tracked up to several hundred milliseconds (Fig. 1C; Movies S1 and S2; SI Appendix). We wondered if the presence of foci was an artifact due to GFP oligomerization. To discourage artifactual aggregation we performed a control using another type of GFP containing an A206K mutation (denoted GFPmut3 or mGFP) known to inhibit oligomerization (22). However, both *in vitro* experiments using purified GFP and mGFP (Fig. S2B; SI Appendix) and live cell experiments at *glucose* (-/+) (Fig. S2E and S2F) indicated no significant difference to foci brightness values (Pearson's $\chi^2$ test, $p<0.001$). We also developed a genomically encoded Mig1 reporter using green-red photoswitchable fluorescent protein mEos2 (23). Super-resolution stochastic optical reconstruction microscopy (STORM) from hundreds of individual photoactivated tracks indicated the presence of foci (SI Appendix), clearly present in nuclei hotspots in live cells at *glucose* (+) (Fig. S1). These results strongly argue that foci formation is not dependent on hypothetical fluorescent protein oligomerization.

We implemented nanoscale tracking based on automated foci detection which combined iterative Gaussian masking and fitting to foci pixel intensity distributions to determine the spatial localization to a lateral precision of 40nm (24,25). Tracking was coupled to stoichiometry analysis using single GFP photobleaching of foci tracks (21) and single cell copy number quantification (26). These methods enabled us to objectively quantify the number of Mig1 molecules associated with each foci, its effective microscopic diffusion coefficient $D$ and spatiotemporal dynamics in regards to its location in the cytoplasm, nucleus or translocating across the nuclear envelope, as well as the copy number of Mig1 molecules associated with each subcellular region and in each cell as a whole. These analyses indicated ~850-1,300 Mig1 total molecules per cell, dependent on extracellular glucose (Fig. 1D; Table S2).

At *glucose* (-) we measured a mean ~950 Mig1 molecules per cell in the cytoplasmic pool (Fig. 1D) and 30-50 multimeric foci in total per cell, based on interpolating the observed number of foci in the microscope's known depth of field over the entirety of the cell volume. These foci had a mean stoichiometry of 6-9 molecules and mean $D$ of 1-2$\mu m^2$/s, extending as high as 6$\mu m^2$/s. In nuclei, the mean foci stoichiometry and $D$ was the same as the cytoplasm to within experimental error (Student $t$-test, $p<0.05$), with a similar concentration. Trans-nuclear foci, those entering /leaving the nucleus during observed tracking, also had the same mean stoichiometry and $D$ to cytoplasmic values to within experimental error ($p<0.05$). However, at *glucose* (+) we measured a considerable increase in the proportion of nuclear foci compared to *glucose* (-), with up to 8 foci per nucleus of mean apparent stoichiometry 24-28 molecules, but $D$ lower by a factor of 2, and 0-3 cytoplasmic/trans-nuclear foci per cell (Fig. 2A and 2B).

**Mig1 cluster localization is dependent on phosphorylation status**

To understand how Mig1 clustering was affected by its phosphorylation we deleted the *SNF1* gene which encodes the Mig1-upstream kinase, Snf1, a key regulator of Mig1 phosphorylation. Under Slimfield imaging this strain indicated Mig1 clusters with similar stoichiometry and $D$ as for the wild type strain at *glucose* (+), but with a significant insensitivity to depleting extracellular glucose (Fig. S1, S3A and S3B). We also used a yeast strain in which the kinase activity of Snf1 could be controllably inhibited biochemically by addition of cell permeable PP1 analog 1NM-PP1. Slimfield imaging indicated similar results in terms of the presence of Mig1 clusters, their stoichiometry and $D$, but again showing a marked insensitivity towards depleted extracellular glucose indistinguishable from the wild type *glucose* (+) phenotype (Fig. S1, S3C, S3D and Table S2). We also tested a strain containing Mig1 with four serine phosphorylation sites (Ser222, 278, 311 and 381) mutated to alanine, which were shown to affect Mig1 localization and phosphorylation dependence on extracellular glucose (27). Slimfield showed the same pattern of localization as the *SNF1* deletion while retaining the presence of Mig1 clusters (Fig. S3E). These results suggest that Mig1 phosphorylation does not affect its ability to form clusters, but does alter their localization bias between nucleus and cytoplasm.

**Cytoplasmic Mig1 is mobile but nuclear Mig1 has mobile and immobile states**

The dynamics of Mig1 between cytoplasm and nucleus is critically important to its role in gene regulation. We therefore interrogated tracked foci mobility. We quantified cumulative distribution functions (CDFs) for all nuclear and cytoplasmic tracks (12). A CDF signifies the probability that foci will move a certain distance from their starting point as a function of time while tracked. A mixed mobility population can be modeled as the weighted sum of multiple CDFs characterized by different $D$ (SI Appendix). Cytoplasmic foci at *glucose* (+/-), and nuclear foci at *glucose* (-), were consistent with just a single mobile population (Fig. S4) whose $D$ of 1-2 $\mu m^2$/s was consistent with earlier observations. However, nuclear foci at *glucose* (+) indicated a mixture of mobile and immobile components (Fig. 3A). These results, substantiated by fitting two Gamma functions to the distribution of estimated $D$ (28) for *glucose* (+) nuclear foci (Fig. 3A, inset), indicate 20-30% of nuclear foci are immobile, consistent with a DNA-bound state. Mean square displacement analysis of foci tracks sorted by stoichiometry indicated Brownian diffusion over short timescales of a few tens of ms but anomalous diffusion over longer timescales >30ms (Fig. 3B; SI Appendix). These results are consistent with *glucose* (+) Mig1 diffusion being restrained by interactions with nuclear structures, similar to that reported for other transcription factors (29). Here however this interaction is dependent on extracellular glucose despite Mig1 requiring a pathway of proteins to detect it, unlike the more direct detection mechanism of the prokaryotic *lac* repressor. A strain in which GFP labeled Mig1 had its Zn finger deleted (17) indicated no significant immobile cluster population at *glucose* (+/-) (Fig. S4). We conclude that Mig1 clusters bind to the DNA via their Zn finger motif with direct glucose dependence.

**Mig1 nuclear translocation selectivity does not depend on glucose but is mediated by interactions away from the nuclear envelope**

Due to the marked localization of Mig1 towards nucleus/cytoplasm at *glucose* (+/-) respectively, we asked whether this spatial bias was due to selectivity initiated during translocation at the nuclear envelope. By converting trans-nuclear tracks into coordinates parallel and perpendicular to the measured nuclear envelope position, and synchronizing origins to be the nuclear envelope crossing point, we could compare spatiotemporal dynamics of different Mig1 clusters during translocation. A heat map of spatial distributions of translocating clusters indicated a hotspot of comparable volume to that of structures of budding yeast nuclear pore complexes (30) and accessory nuclear structures of cytoplasmic nucleoporin filaments and nuclear basket (31), with some nuclear impairment to mobility consistent with restrained mobility (Fig. 3C). We observed a dwell in cluster translocation across the 30-40nm width of the nuclear envelope (Fig. 3D). At *glucose* (+) the proportion of detected trans-nuclear foci was significantly higher compared to *glucose* (-), consistent with Mig1's role to repress genes. The distribution of dwell times could be fitted using a single exponential function with ~10ms time constant similar to previous estimates for transport factors (32). However, although the relative proportion of trans-nuclear foci was much lower at *glucose* (-) compared to *glucose* (+), the dwell time constant was found to be insensitive to glucose (Fig. 3E). This insensitivity to extracellular chemical signal demonstrates, surprisingly, that there is no direct selectivity on the basis of transcription factor phosphorylation state by nuclear pore complexes themselves, suggesting that cargo selectivity mechanisms of nuclear transport (33), as reported for a range of substrates, is blind to the phosphorylation state. Coupled with our observation that Mig1 at *glucose* (-) does not exhibit immobility in the nucleus, this suggests that Mig1 localization is driven by changes in Mig1 binding affinity to the DNA or to other proteins within or outside the nucleus not involving the nuclear pore complex.

**Mig1 nuclear foci bound to targets turn over slowly as whole clusters in >100s**

To further understand the mechanisms of Mig1 binding/release during gene regulation we sought to quantify kinetics of these events at Mig1 targets. By modifying our microscope we could implement an independent focused laser path using the same laser source, enabling us to use fluorescence recovery after photobleaching (FRAP) to probe nuclear Mig1 turnover (SI Appendix). The focused laser rapidly photobleached GFP content in cell nuclei in <200ms (Fig. 3F). We could then monitor recovery of any fluorescence intensity by illuminating with millisecond Slimfield stroboscopically as opposed to continuously to extend the observation timescale to >1,000s. Using automated foci detection we could separate nuclear pool and foci content at each time point for each cell. These analyses demonstrated measurable fluorescence recovery for both components, which could be fitted by single exponentials indicating fast recovery of pool at both *glucose* (-) and (+) with a time constant <5s but a larger time constant at *glucose* (+) for nuclear foci >100s (Fig. 3G). Further analysis of intensity levels at each time point revealed a stoichiometry periodicity in nuclear foci recovery equivalent to 7-9 GFP molecules (Fig. S5A), but no obvious periodicity in stoichiometry measurable from pool recovery. An identical periodicity within experimental error was measured from nuclear foci at *glucose* (+) in steady-state (Fig. 4A). These periodicity values in Mig1 stoichiometry were consistent with earlier observations for cytoplasmic and trans-nuclear clusters at *glucose* (+/-), and in the nucleus at *glucose* (-), with mean stoichiometry ~7 molecules. These data taken as a whole clearly suggest that molecular turnover at nuclear foci of Mig1 bound to its target genes occurs in units of single clusters, as opposed to single Mig1 monomers.

**Mig1 clusters are spherical, a few tens of nm wide**

Our observations from stoichiometry, dynamics and kinetics, which supported the hypothesis that functional clusters of Mig1 perform the role of gene regulation, also suggested an obvious prediction in terms of the size of observed foci: the physical diameter of a multimeric cluster should be larger than that of a single Mig1 monomer. We therefore sought to quantify foci widths from Slimfield data by performing intensity profile analysis on background-corrected pixel values over each foci image (SI Appendix). The diameter was estimated from the measured width corrected for motion blur due to particle diffusion in the sampling time of a single image frame, minus that measured from single purified GFP molecules immobilized to the coverslip surface in separate *in vitro* experiments. This analysis revealed diameters of 15-50nm at *glucose* (-), which showed an increase with foci stoichiometry $S$ that could be fitted with a power law dependence $S^a$ (Fig. S5B) with optimized exponent $a$ of $0.32 \pm 0.06$ (±SEM). Immuno-gold electron microscopy of fixed cells probed with anti-GFP antibody confirmed the presence of GFP in 90nm cryosections with some evidence of clusters containing up to 7 Mig1 molecules (Fig. S5C). A heuristic tight packing model for GFP labeled Mig1 monomers in each cluster predicts that, in the instance of an idealized spherical cluster, $a = 1/3$. Our data at *glucose* (-) thus supports the hypothesis that Mig1 clusters have a spherical shape. For nuclear foci at *glucose* (+) we measured larger apparent diameters and stoichiometries, consistent with >1 individual Mig1 cluster being separated by less than our measured ~200nm optical resolution limit. This observation agrees with earlier measurements of stoichiometry periodicity for nuclear foci at *glucose* (+). In other words, that higher apparent stoichiometry nuclear foci are consistent with multiple individual Mig1 clusters each containing ~7 molecules separated by a nearest neighbor distance <200nm and so detected as a single fluorescent foci.

**Clusters are stabilized by depletion forces**

Since we observed Mig1 clusters in live cells using Slimfield imaging we wondered if these could be detected and further quantified using other methods. However, native gel electrophoresis on extracts from Mig1-GFP cells (Fig. S5D) indicated a single stained band for Mig1, which was consistent with denaturing SDS-PAGE combined with western blotting using recombinant Mig1-GFP, and protein extracts from the parental cells which included no fluorescent reporter (Fig. S5E and S5F). Slimfield imaging on purified Mig1-GFP *in vitro* under identical imaging conditions for live cells similarly indicated monomeric Mig1-GFP foci in addition to a small fraction of brighter foci which were consistent with predicted random overlap of monomer images (SI Appendix). However, on addition of low molecular weight polyethylene glycol (PEG) at a concentration known to mimic small molecule 'depletion' forces in live cells (34) we detected significant numbers of multimeric foci (Fig. 4B). Depletion is an entropic derived attractive force which results from osmotic pressure between particles suspended in solution that are separated by distances short enough to exclude other surrounding smaller particles. Purified GFP alone under identical conditions showed no such effect (Fig. S5G). These results support a hypothesis that clusters present in live cells are stabilized by depletion components that are lost during biochemical purification.

**Chromosome structure modeling supports a cluster binding hypothesis**

We speculated that Mig1 cluster-mediated gene regulation had testable predictions in regards to the nuclear location of Mig1 at elevated extracellular glucose. We therefore developed quantitative models to simulate the appearance of realistic images of genome-bound Mig1-GFP at *glucose* (+). We used sequence analysis to infer locations of Mig1 binding sites in the yeast genome (SI Appendix), based on alignment matches to previously identified 17bp Mig1 target patterns (35) which comprised conserved AT-rich 5bp and GC-rich 6bp sequences. In scanning the entire *S. cerevisiae* genome we found >3,000 hits though only 112 matches for likely gene regulatory sites located in promoter regions (Table S3). We mapped these candidate binding sites onto specific 3D locations (Fig. 4C) obtained from a consensus structure for budding yeast chromosomes based on 3C data (36). We generated simulated images, adding experimentally realistic levels of signal and noise, and ran these synthetic data through the same tracking software as for experimental data. We used identical algorithm parameters throughout and compared these predictions to the measured experimental stoichiometry distributions.

In the first instance we used these locations as coordinates for Mig1 monomer binding, assuming that just a single Mig1 molecule binds to a target. Copy number analysis of Slimfield data (Table S2) indicated a mean ~190 Mig1 molecules per cell associated with nuclear foci, greater than the number of Mig1 binding sites in promoter regions. We assigned 112 molecules to target promoter binding sites, then assigned the remaining 78 molecules randomly to non-specific DNA coordinates of the chromosomal structure. We included the effects of different orientations of the chromosomal structure relative to the camera by generating simulations from different projections and included these in compiled synthetic datasets.

We then contrasted monomer binding to a cluster binding model, which assumed that a whole cluster comprising 7 GFP labeled Mig1 molecules binds a single Mig1 target. Here we randomly assigned the 190 Mig1 molecules into just 27 (i.e. ~190/7) 7-mer clusters to the set of 112 Mig1 target promoter sites. We also implemented improvements of both monomer and cluster binding models to account for the presence of trans-nuclear tracks. Extrapolating the number of detected trans-nuclear foci in our microscope's depth of field over the whole nuclear surface area indicated a total of ~130 Mig1 molecules at *glucose* (+) inside the nucleus prior to export across the cytoplasm. We simulated the presence of these trans-nuclear molecules either using 130 GFP-labeled Mig1 molecules as monomers, or as 18 (i.e. ~130/7) 7-mer clusters at random 3D coordinates over the nuclear envelope surface (Fig. S5H).

We discovered that a cluster binding model which included the presence of trans-nuclear foci generated excellent agreement to the experimental foci stoichiometry distribution ($R^2$=0.75) compared to a very poor fit for a monomer binding model ($R^2$<0) (Fig. 4D). The optimized cluster model fit involved on average ~25% of promoter loci to be bound across a population of simulated cells by a 7-mer cluster with the remaining clusters located non-specifically, near the nuclear envelope, consistent with nuclear transit. This structural model supports the hypothesis that the functional unit of Mig1-mediated gene regulation is a cluster of Mig1 molecules, as opposed to Mig1 acting as a monomer.

**The activator Msn2 also forms functional clusters**

We wondered if the discovery of transcription factor clusters was unique to specific properties of the Mig1 repressor, as opposed to being a more general feature of other Zn finger transcription factors. To address this question we prepared a genomically encoded GFP fusion construct of a similar protein Msn2. Nrd1-mCherry was again used as a nuclear marker (Fig. S1). Msn2 acts as an activator and not a repressor, which co-regulates several Mig1 target genes but with the opposite nuclear localization response to glucose (17). On performing Slimfield under identical conditions to the Mig1-GFP strain we again observed a significant population of fluorescent Msn2 foci, which had comparable $D$ and stoichiometry to those estimated earlier for Mig1 (Table S2). The key difference with the data from the Mig1-GFP strain was that Msn2, unlike Mig1, demonstrated high apparent foci stoichiometry values and lower values of $D$ at *glucose* (-), which was consistent with its role as an activator of the same target genes as opposed to a repressor (Fig. S6A and S6B). Immuno-gold electron microscopy of fixed Msn2-GFP cells confirmed the presence of GFP in 90nm cryosections with evidence for clusters of comparable diameters to Mig1-GFP (Fig. S5C). These results suggest that two different eukaryotic transcription factors that have antagonist effects on the same target genes operate as molecular clusters.

To test the functional relevance of Mig1 and Msn2 clusters we performed Slimfield on a strain in which Mig1 and Msn2 were genomically labeled using mCherry and orange fluorescent protein mKO2, respectively (17). This strain also contained a plasmid with GFP labeled PP7 protein to report on nuclear mRNA expressed specifically from the glycogen synthase*GSY1* gene, whose expression can be induced by glucose starvation and is a target of Mig1 and Msn2, labelled with 24 repeats of the PP7 binding sequence (37). In switching from *glucose* (+) to (-) and observing the same cell throughout, we measured PP7 accumulating with similar localization patterns to those of Mig1 clusters at *glucose* (+) (Fig. S6C). We calculated the numerical overlap integral between these Mig1 and PP7 foci (Fig. S6D), indicating a high mean of ~0.95, where 1 is the theoretical maximum for 100% colocalization in the absence of noise (25). We also observed similar high colocalization between Msn2-mKO2 clusters and PP7-GFP at *glucose* (-) (Fig. S6E). These results demonstrate a functional link between the localization of Mig1 and Msn2 clusters, and the transcribed mRNA from their target genes.

## Mig1 and Msn2 possess intrinsic disorder which may favor clustering

Since both Mig1 and Msn2 demonstrate significant populations of clustered molecules in functional cell strains we asked the question if there were features common to the sequences of both proteins which might explain this behavior. To address this question we used multiple sequence alignment to determine conserved structural features of both proteins, and secondary structure prediction tools with disorder prediction algorithms. As expected, sequence alignment indicated the presence of the Zn finger motif in both proteins, with secondary structure predictions suggesting relatively elongated structures (Fig. 5A). However, disorder predictions indicated multiple extended intrinsically disordered regions in both Mig1 and Msn2 sequences with an overall proportion of disordered content >50%, as high as 75% for Mig1 (Fig. 5B; Table S4). We measured a trend from a more structured region of Mig1 towards the N-terminus and more disordered regions towards the C-terminus. Msn2 demonstrated a similar bipolar trend but with the structured Zn finger motif towards the C-terminus and the disordered sequences towards the N-terminus. We then ran the same analysis as a comparison against the prokaryotic transcription factor LacI, which represses expression from genes of the *lac* operon as part of the prokaryotic glucose sensing pathway. The predicted disorder content in the case of LacI was <50%. In addition, further sequence alignment analysis predicted that at least 50% of candidate phosphorylation sites in either Mig1 or Msn2 lie within these intrinsically disordered sequences (Table S4; Fig. 5A). An important observation reported previously is that the comparatively highly structured LacI exhibits no obvious clustering behavior from similar high-speed fluorescence microscopy tracking on live bacteria (4). Intrinsically disordered proteins are known to undergo phase transitions which may enable cluster formation and increase the likelihood of binding to nucleic acids (38,39). We measured significant changes in circular dichroism (SI Appendix) of the Mig1 fusion construct upon addition of PEG in the wavelength range 200-230nm (Fig. 5C) known to be sensitive to transitions between ordered and intrinsically disordered states (40,41). Since the Zn finger motif lies towards the opposite terminus to the disordered content for both Mig1 and Msn2 this may suggest a molecular bipolarity which could stabilize a cluster core while exposing Zn fingers on the surface enabling interaction with accessible DNA. This structural mechanism has analogies to that of phospholipid interactions driving micelle formation, however mediated here through disordered sequence interactions as opposed to hydrophobic forces (Fig. 5C). The prevalence of phosphorylation sites located in disordered regions may also suggest a role in mediating affinity to target genes, similar to protein-protein binding by phosphorylation and intrinsic disorder coupling (42).

## Discussion

Our findings address a totally underexplored and novel aspect of gene regulation with technology that has not been available until recently. In summary, we observe that the repressor protein Mig1 forms clusters which, upon extracellular glucose detection, localize dynamically from the cytoplasm to bind to locations consistent withpromoter sequences of its target genes. . Similar localization events were observed for the activator Msn2 under glucose limiting conditions. Moreover, Mig1 and Msn2 oligomers colocalized with mRNA transcribed from *GSY1* gene at glucose (+/-), respectively. Our results therefore strongly support a functional link between Mig1 and Msn2 transcription factor clusters and target gene expression. The physiological role of multivalent transcription factor clusters has been elucidated through simulations (9) but unobserved until now. These simulations show that intersegmental transfer between sections of nuclear DNA was essential for factors to find their binding sites within physiologically relevant timescales and requires multivalency. Our findings address the longstanding question of how transcription factors find their targets in the genome so efficiently. Evidence for higher molecular weight Mig1 states from biochemical studies has been suggested previously (43). A Mig1-His-HA construct was overexpressed in yeast and cell extracts run in different glucose concentrations through sucrose density centrifugation. In western blots, a higher molecular weight band was observed, attributed to a hypothetical cofactor protein. However, no cofactor was detected and none reported to date. The modal molecular weight observed was ~four times that of Mig1 but with a wide observed distribution consistent with our mean detected cluster size of ~7 molecules. The authors only reported detecting higher molecular weight states in the nucleus in repressing conditions.

Our measured turnover of genome-bound Mig1 has similar timescales to that estimated for nucleoid-bound LacI (4), but similar rates of turnover have also been observed in yeast for a DNA-bound activator (44). Faster off rates have been observed during single particle tracking of the DNA-bound fraction of the glucocorticoid receptor (GR) transcription factor in mammalian cells, equivalent to a residence time on DNA of just 1s (12). Single GR molecules appear to bind as a homodimer complex on DNA, and slower Mig1 off rates may suggest higher order multivalency, consistent with Mig1 clusters.

Estimating nearest-neighbor distances between Mig1 promoter sites in the *S. cerevisiae* genome from the 3C model (Fig. 5D) indicates 20-30% are <50 nm, small enough to enable different DNA segments to be linked though intersegment transfer by a single cluster (6,9). This separation would also enable simultaneous binding of >1 target (Fig. 5E). The proportion of loci separated by <50nm is also consistent with the estimated proportion of immobile foci and with the proportion of cluster-occupied sites predicted from our structural model. Such multivalency chimes with the tetrameric binding of prokaryotic LacI leading to similar low promoter off rates (4). Extensive bioinformatics analysis of proteome disorder across a range of species suggests a sharp increase from prokaryotes to eukaryotes (45), speculatively due to the prokaryotic absence of cell compartments and regulated ubiquitination mechanisms lowering protection of unfolded disordered structures from degradation (46). Our discovery in yeast may reveal a eukaryotic adaptation that stabilizes gene expression. The slow off rate we measure would result in insensitivity to high frequency stochastic noise which could otherwise result in false positive detection and an associated wasteful expression response. We also note that long turnover times may facilitate modulation between co-regulatory factors by maximizing overlap periods, as suggested previously for Mig1/Msn2 (17).

Our results suggest that cellular depletion forces due to crowding enable cluster formation. Crowding is known to increase oligomerization reaction rates for low association proteins but slow down fast reactions due to an associated decrease in diffusion rates, and have a more pronounced effect on higher order multimers rather than dimers (34). It is technically challenging to study depletion forces *in vivo*, however there is growing *in vitro* and *in silico* evidence of the importance of molecular crowding in cell biology. A particularly striking effect was observed previously in the formation of clusters of the bacterial cell division protein FtsZ in the presence of two crowding proteins – hemoglobin and BSA (47). Higher order decamers and multimers were observed in the presence of crowding agents and these structures are thought to account for as much as 1/3 of the *in vivo* FtsZ content. Similarly, two recent yeast studies of the high-osmolarity glycerol (HOG) pathway also suggest a dependence on gene expression mediated by molecular crowding (48,49).

The range of GFP labeled Mig1 cluster diameters *in vivo* of 15-50nm is smaller than the 80nm diameter of yeast nuclear pore complexes (50), not prohibitively large as to prevent intact clusters from translocating across the nuclear envelope. An earlier *in vitro* study using sucrose gradient centrifugation suggested a Stokes radius of 4.8 nm for the Mig1 fraction, i.e. diameter 9.6nm, large for a Mig1 monomer (43) whose molecular weight is 55.5kDa, e.g. that of monomeric bovine serum albumin (BSA) at a molecular weight of 66kDa is closer to 3.5nm (51). The authors ascribed this effect to a hypothetical elongated monomeric structure for Mig1. The equivalent Stokes radius for GFP has been measured at 2.4nm (52), i.e. diameter 4.8nm. Also, for our Mig1-GFP construct there are two amino acids residues in the linker region between the Mig1 and GFP sequences (i.e. additional length 0.7-0.8nm). Thus the anticipated hydrodynamic diameter of Mig1-GFP is 15-16nm. The mean observed ~7-mer cluster diameter from Slimfield data is ~30nm, which, assuming a spherical packing geometry, suggests a subunit diameter for single Mig1-GFP molecules of ~$30/7^{1/3} \approx 15.6$nm, consistent with that predicted from the earlier hydrodynamic expectations. Using Stokes law this estimated hydrodynamic radius indicates an effective viscosity for the cytoplasm and nucleoplasm as low as 2-3cP, compatible with earlier live cell estimates on mammalian cells using fluorescence correlation spectroscopy (FCS) (53).

One alternative hypothesis to that of intrinsically disordered sequences mediating Mig1 cluster formation is the existence of a hypothetical cofactor protein to Mig1. However, such a cofactor would be invisible on our Slimfield assay but would result in a larger measured hydrodynamic radius than we estimate from fluorescence imaging, which would be manifest as larger apparent viscosity values than those we observe. Coupled to observations of Msn2 forming clusters also, and the lack of any reported stable cofactor candidate to date, limits the cofactor hypothesis. Pull down assays do suggest that promoter bound Mig1 consists of a complex which includes the accessory proteins Ssn6 and Tup1 (54), however this would not explain the observation of Mig1 clusters outside the nucleus.

There may be other advantages in having a different strategy between *S. cerevisiae* and *E. coli* to achieve lowered transcriptional regulator off rate. A clue to these may lie in phosphorylation. We discovered that at least 50% of candidate serine or threonine phosphorylation sites in Mig1 and Msn2 lie in regions with high intrinsic disorder, which may have higher sequence-unspecific binding affinities to DNA (38,39). Thus phosphorylation at sites within these regions may potentially disrupt binding to DNA, similar to observed changes to protein-protein affinity being coupled to protein phosphorylation state (42). Previous studies indicate that dephosphorylated Mig1 binds to its targets (55). Thus, intrinsic disorder may be required for bistability in affinity of Mig1/Msn2 to DNA.

Wide scale bioinformatics screening reveals a significant prevalence of intrinsic disorder in eukaryotic transcription factors (56). Our discovery is the first, to our knowledge, to make a link between predicted disorder and the ability to form higher-order clusters in transcription factors. Thus, our results address the longstanding question of why there is so much predicted disorder in eukaryote transcription factors. Our observations that protein interactions based on weak intracellular forces and molecular crowding has direct functional relevance may stimulate new research lines in several areas of cell biology. For example, our findings may have important mechanistic implications for other aggregation processes mediated through intrinsic disorder interactions, such as those of amyloid plaques found in neurodegenerative disorders including Alzheimer's and Parkinson's diseases (57). Increased understanding of the clustering mechanism may not only be of value in understanding such diseases, but could enable future novel synthetic biology applications to manufacture gene circuits with, for example, a range of bespoke response times.

## Materials and methods

### Strain construction

We developed Mig1 fluorescent protein strains based on strain YSH1351 (16) using eGFP in the first instance and also mGFP/GFPmut3 designed to inhibit oligomerization (22), and photoswitchable mEos2 (23), detailed in SI Appendix.

### Single-molecule imaging

A dual-color bespoke laser excitation single-molecule fluorescence microscope was used (20,26) utilizing narrow epifluorescence excitation of 10μm full width at half maximum (FWHM) in the sample plane to generate Slimfield illumination, detailed in SI Appendix.

### Foci tracking and copy number analysis

Foci were automatically detected using software written in MATLAB (Mathworks) (24), lateral localization ~40nm, enabling estimates of *D* and stoichiometry. Copy umbers for individual cells were estimated using image convolution (26). Full details in SI Appendix.

### Structural, mobility and bioinformatics analysis

Circular dichroism was performed on purified GFP labeled Mig1 on a Jasco J810 circular dichromator in sodium phosphate buffer supplemented with 1kDa PEG as appropriate. Transmission electron microscopy was performed on a 120kV Tecnai 12 BioTWIN (FEI) with SIS Megaview III camera on 90nm cryosections of Mig1-GFP or Msn2-GFP fixed cells and probed with anti-GFP and gold-tagged antibodies. The mobility of tracked particles was analyzed using multiple CDFs and Gamma function fits to the diffusion coefficient probability density functions. Multiple sequence alignment was performed using PSI-BLAST with intrinsic disorder search algorithms DISOPRED and PONDR, and PyMOL secondary structure prediction. Full details in SI Appendix.

### Acknowledgments

Supported by the Biological Physical Sciences Institute, Royal Society, MRC (grant MR/K01580X/1), BBSRC (grant BB/N006453/1), Swedish Research Council and European Commission via Marie Curie-Network for Initial training ISOLATE (Grant agreement nr: 289995). We thank Magnus Alm Rosenblad and Sarah Shammas for assistance with RNABOB and PONDR, Marija Cvijovic and Michael Law for help with qPCR data analysis, Andrew Leech and Meg Stark for help with CD and EM. Thanks to Mark Johnston (CU Denver) for donation of Mig1 phosphorylation mutant plasmid, and Michael Elowitz (Caltech) for donation of the Mig1/Msn2/PP7 and Zn finger deletion strain.

### Conflict of interest
All the authors declare that they have no conflict of interests.

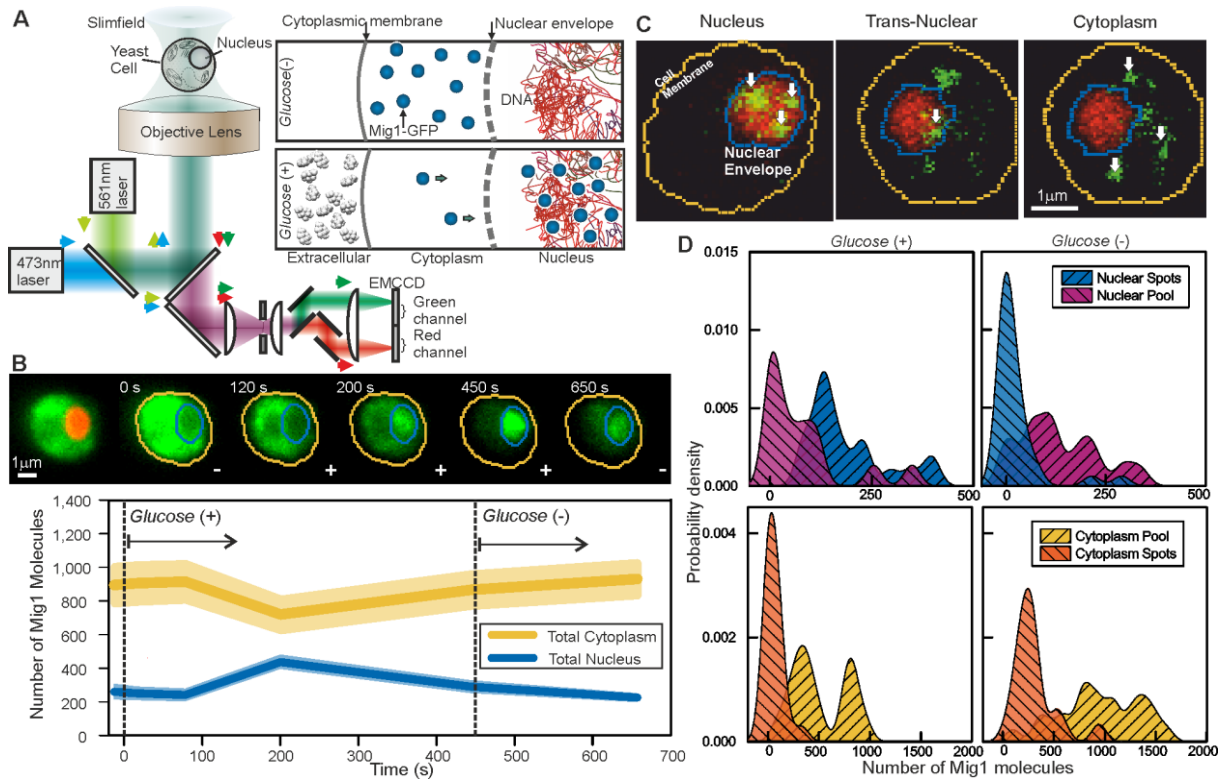

**Fig. 1. Single-molecule Slimfield microscopy of live cells reveals Mig1 clusters.** (**A**) Dual-color fluorescence microscopy assay. Mig1-GFP localization change (cyan, right panels) depending on glucose availability. (**B**) Example of change of Mig1-GFP localization with glucose for the same cell, nuclear Nrd1-mCherry indicated (red, left), mean and SEM errorbounds of total cytoplasmic (yellow) and nuclear (blue) contributions shown (lower panel), n=15 cells. (**C**) Example cells showing nuclear (left), trans-nuclear (center) and cytoplasmic (right) Mig1-GFP localization (green, distinct foci white arrows), Nrd1-mCherry (red) and segmented cell body (yellow) and nuclear envelope (blue) indicated. (**D**) Kernel density estimations for Mig1-GFP content in pool and foci for cytoplasm and nucleus at *glucose* (+/-), n=30 cells.

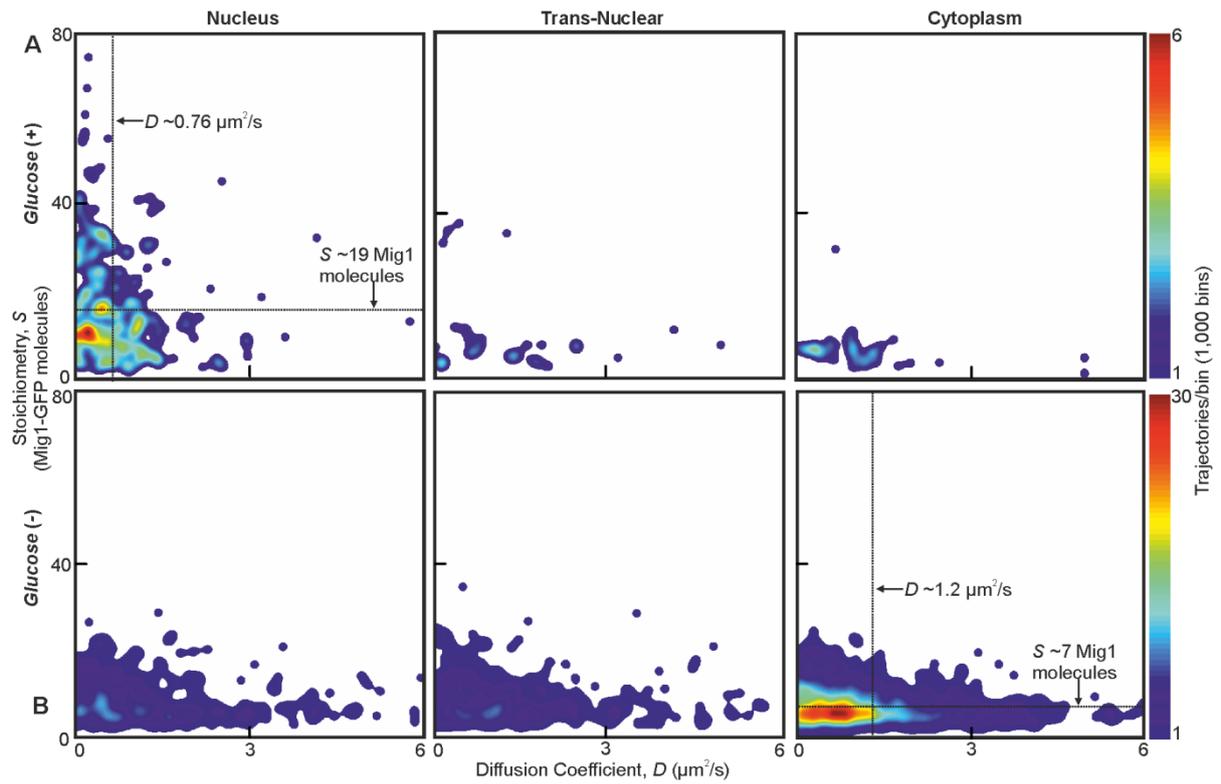

**Fig. 2. Mig1 foci stoichiometry, mobility and localization depend on glucose.** Heat map showing dependence of stoichiometry of detected GFP-labeled Mig1 foci with $D$ under (**A**) *glucose* (+) and (**B**) *glucose* (-) extracellular conditions. Mean values for *glucose* (+) nuclear and *glucose* (-) cytoplasmic foci indicated (arrows).

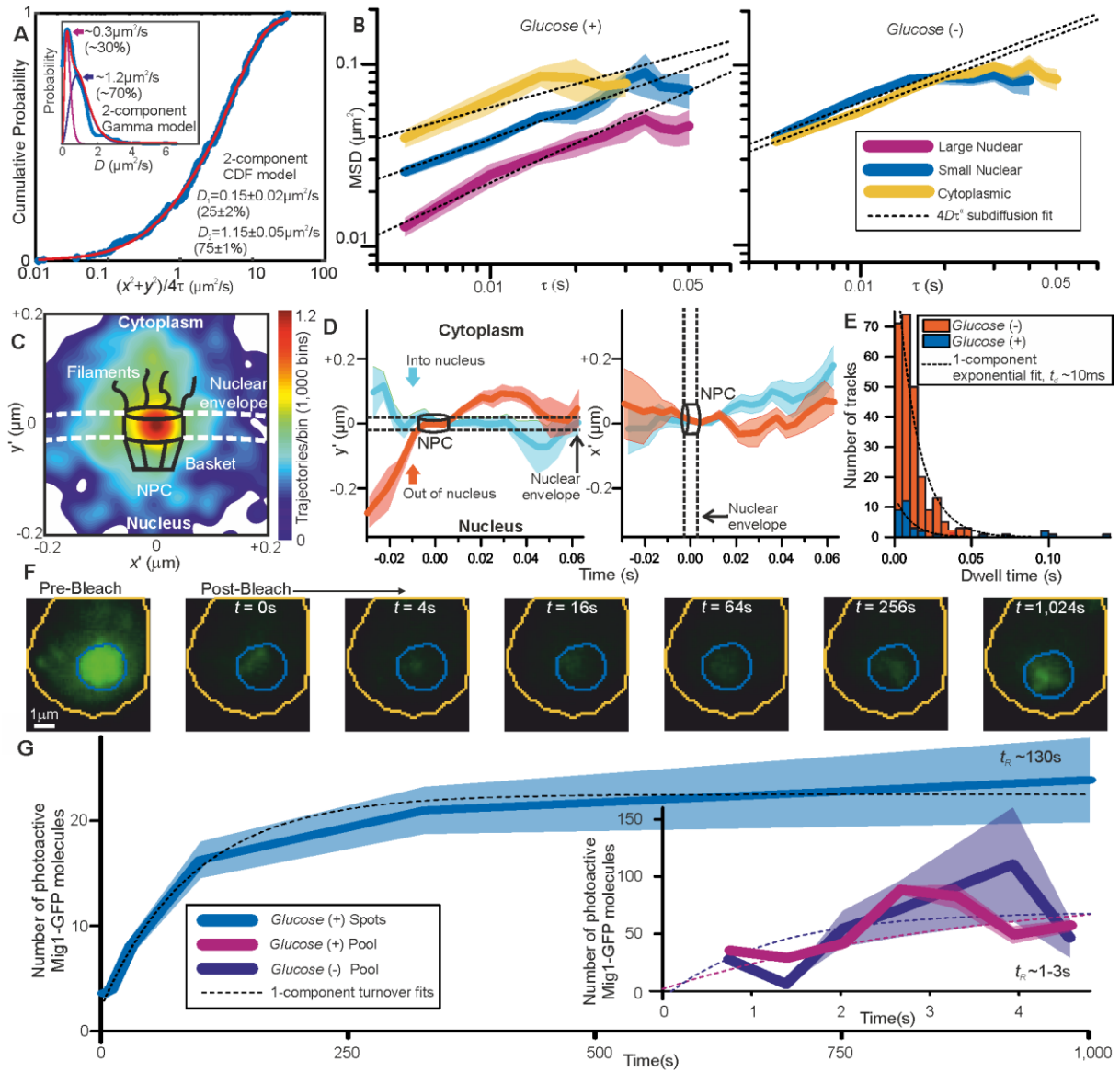

**Fig. 3. Repressor clusters have heterogeneous mobility depending on localization.** (**A**) Cumulative probability, *glucose* (+) nuclear tracks, dual Gamma fit to *D* (inset). (**B**) Mean MSD from cytoplasmic (yellow), small (blue, stoichiometry ≤20 Mig1-GFP molecules) and large nuclear (purple, stoichiometry >20 Mig1-GFP molecules) foci, SEM indicated, n=30 cells. Subdiffusion fits to time intervals ≤30ms (dashed), anomalous coefficient α=0.4-0.8. (**C**) Heat map for trans-nuclear tracks, (**D**) distance perpendicular and parallel to nuclear envelope with time, (**E**) dwell times and single exponential fits (dotted). (**F**) Example *glucose* (+) single cell FRAP, (**G**) mean and SEM indicated, n=5 and 7 cells for *glucose* (-/+) respectively.

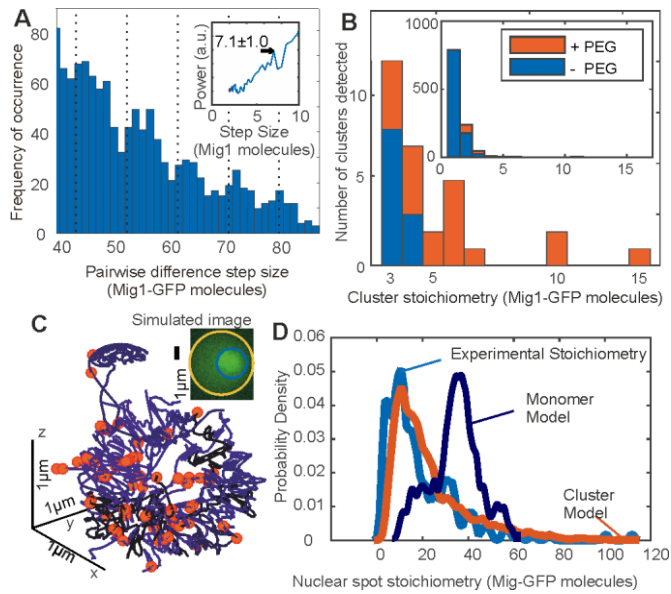

**Fig. 4. Mig1 clusters are stabilized by depletion forces and bind to promoter targets. (A)** Zoom-in on pairwise difference distribution for stoichiometry of Mig1-GFP foci, 7-mer intervals (dashed) and power spectrum (inset), mean and Gaussian sigma error (arrow). **(B)** Stoichiometry for Mig1-GFP clusters *in vitro* in PEG absence (blue)/presence (red). **(C)** 3C model (blue) with overlaid Mig1 promoter binding sites from bioinformatics (red), simulated image based on model with realistic signal and noise added (inset). **(D)** Cluster (red) and monomer (dark blue) model (goodness-of-fit $R^2<0$) for Mig1-GFP stoichiometry compared against experimental data (cyan, $R^2=0.75$).

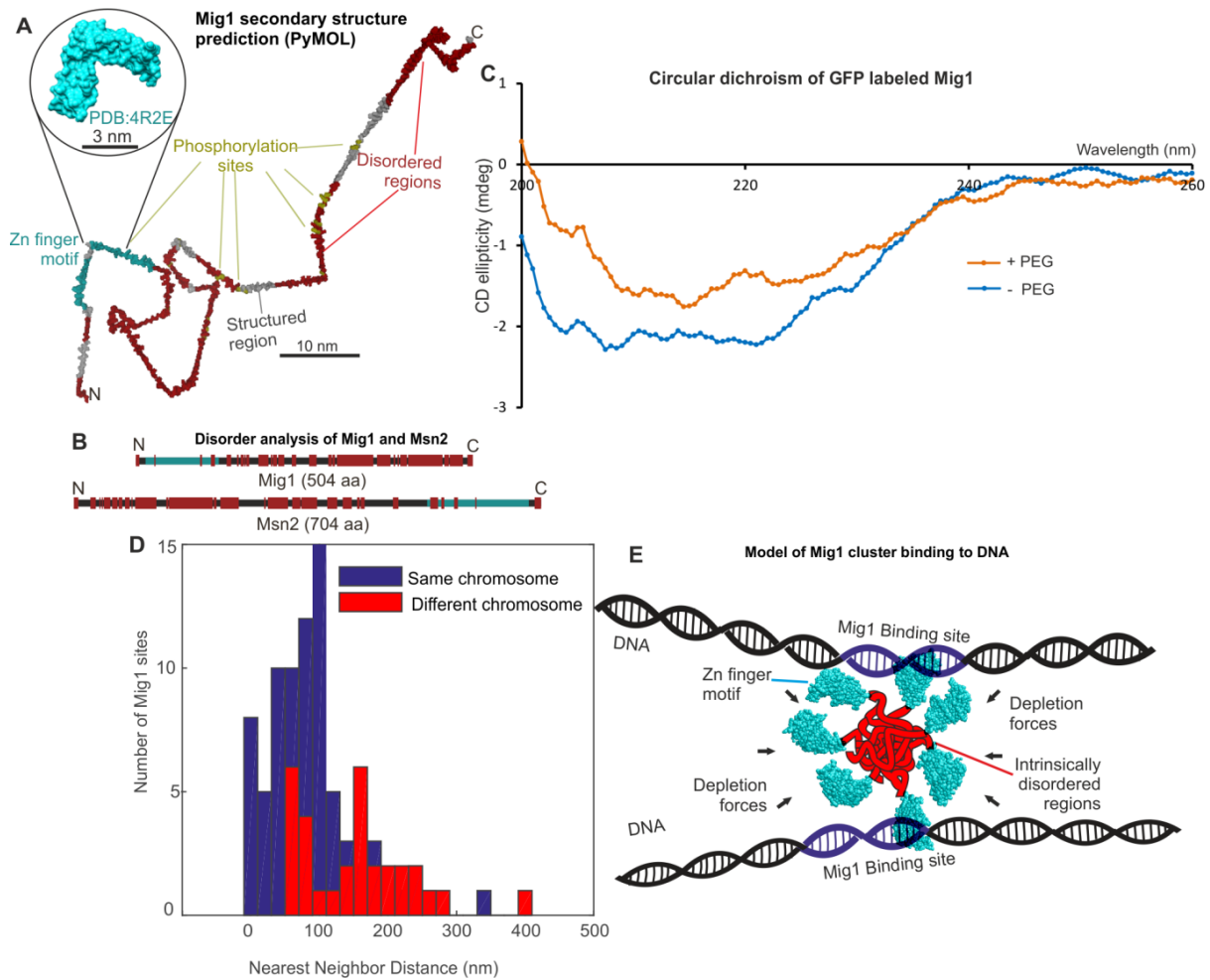

**Fig. 5. Mig1 and Msn2 contain disordered sequences which may mediate cluster formation.** (**A**) Structural prediction for Mig1; Zn finger motif (cyan), disordered sections (red) from PyMOL, beta sheet (gray), phosphorylation sites (yellow); zoom-in indicates structure of conserved Zn finger from PSI-BLAST to PDB ID: 4R2E (Wilms tumor protein, WT1). (**B**) DISOPRED prediction for Mig1 and Msn2; disordered regions (red), Zn finger regions (cyan). (**C**) Circular dichroism of Mig1-GFP *in vitro* in PEG absence (blue)/presence (orange) (**D**) Distribution of nearest neighbor distances for Mig1 sites within promoters on same (blue) or different (red) chromosome. (**E**) Schematic of depletion-stabilized Mig1 cluster bound to multiple promoter targets (Zn finger PDB ID: 4R2E).

# Supporting Information for

# Transcription factor clusters regulate genes in eukaryotic cells

**Authors:** Adam J. M. Wollman[a,1], Sviatlana Shashkova[a,b,1], Erik G. Hedlund[a], Rosmarie Friemann[b], Stefan Hohmann[b,c], Mark C. Leake[a,2]

**Affiliations:**
[a] Biological Physical Sciences Institute, University of York, York YO10 5DD, UK

[b] Department of Chemistry and Molecular Biology, University of Gothenburg, 40530 Göteborg, Sweden.

[c] Department of Biology and Biological Engineering, Chalmers University of Technology, 41296 Göteborg, Sweden.

[1] A.J.M.W. and S.S. contributed equally to this work.

[2] To whom correspondence should be addressed. Email: mark.leake@york.ac.uk.

**This PDF includes:**
Supplementary Materials and methods
Table S1 *S. cerevisiae* cell strains and plasmids
Table S2. Foci tracking data
Table S3. Number of potential Mig1 target promoter sites per chromosome
Table S4. Bioinformatics analysis for intrinsically disordered sequences.
Supplementary captions and stills Movies S1 and S2
Fig. S1. Brightfield and fluorescence micrographs of key strains and glucose conditions.
Fig. S2. Fluorescent reporter strains have similar viability to wild type, with relatively fast maturation of fluorescent protein, and no evidence for GFP-mediated oligomerization.
Fig. S3. Cumulative probability distance analysis reveals a single mobile population in the cytoplasm at glucose (+/-) and in the nucleus and glucose (-).
Fig. S4. Mig1 nuclear foci turnover in clusters whose diameter is a few tens of nm.
Fig. S5. GFP-labeled Mig1 clusters are not formed due to multimerization of GFP.
Fig. S6. Simulations based on 3D chromosomal structural modeling support the hypothesis that Mig1 translocates as whole clusters across the nuclear envelope and binds as whole clusters to promoter targets.

## Materials and methods

### Strain construction and characterization

Mig1-mGFP and Mig1-mEos2 fusions were constructed by introducing into YSH1351 (BY4741 wild type) cells the *mGFP-HIS3* or *mEOs2-HIS3* PCR fragment flanked on its 5' end with 50bp sequence of *MIG1* 3' end and 50bp downstream of *MIG1* excluding the STOP codon. The *mEOs2-HIS3* and *mGFP-HIS3* fragment was amplified from mEOs-his plasmid (GeneArt, Life Technologies) and pmGFP-S plasmid designed for this study by inserting into plasmid YDp-H. Modified strains in which the *SNF1* gene was deleted, *snf1Δ*, were prepared by compromising the gene with an auxotrophic marker providing into cells the *LEU2* PCR product amplified from plasmid YDp-L and flanked with 50bp of *SNF1* upstream and downstream sequence on 5' and 3' ends, respectively. Strains in which Snf1 kinase activity can be inhibited by 25µM 1NM-PP1 were prepared by introducing into cells a plasmid with an ATP analog-sensitive version of Snf1 with *I132G* mutation (58). All transformations were performed using the lithium-acetate protocol (59).

Cell doubling times of all strains were calculated (60) (fig. S2) based on $OD_{600}$ values obtained during cultivation (Bioscreen analyser C). We quantified mRNA relative expression of the *MIG1* gene using qPCR against the constitutive actin gene *ACT1* in the wild type and the Mig1-mGFP strain in cells pre-grown in 4% glucose and then shifted to elevated (4%) and depleted (0.2%) extracellular glucose for 2 h. mRNA isolation and cDNA synthesis were performed as described previously (61).

For Msn2-GFP experiments we used the YSH2350 strain (*MATa msn2-GFP-HIS3 nrd1-mCherry-hphNT1 MET LYS*) in BY4741 background.

### Protein production and purification

His-tagged *mCherry*, *eGFP* and *mGFP* genes were amplified by PCR and cloned into pET vectors. An expression pRSET A plasmid containing 6xHis-Mig1-mGFP was obtained commercially (GeneArt, Life Technologies). *Escherichia coli* strain BL21(DE3) carrying the expression plasmid was grown in LB with 100µg/ml ampicillin and 34µg/ml chloramphenicol at 37ºC to $OD_{600}$ 0.7. Protein expression was induced by adding isopropyl-β-D-thiogalactopyranoside (IPTG) at final concentration of 1mM for 3h at 30°C. Cells were suspended in 50mM $NaH_2PO_4$, 10mM Tris, 300mM NaCl, 2mM EDTA, 0.2mM PMSF, 0.1% β-mercaptoethanol, pH 8.0, and lysed by sonication or by three passages through a chilled Emulsiflex (Avestin). Extracts were cleared (24,000g, 30min) and filtered (pore diameter 0.45µm; Millipore, Bedford). All proteins were purified using $Ni^{2+}$ affinity chromatography on a 5ml HisTrap FF column (GE Healthcare). Mig1-mGFP was eluted with a linear gradient 0 - 0.4 M imidazole in lysis buffer. Mig1-mGFP was further purified by size-exclusion chromatography (Superdex 200 Increase 10/300, GE Healthcare) and concentrated (50 kDa molecular weight cutoff VIVASPIN 20 concentrator). Purity of the sample was confirmed by Coomassie stained SDS-PAGE gels (Simply Blue Safe Stain, Life Technologies).

### Media and growth conditions

Cells from frozen stocks were grown on plates with standard YPD media (10 g/l yeast extract, 20 g/l bacto-peptone, 20 g/l agar) supplemented with 4% glucose (w/v) at $30^0$C overnight. For the liquid cultures, the YPD was prepared as above but without agar, and the cells were grown at $30^0$C while shaking (180 rpm).

For transformants that carried a plasmid with mutated *SNF1* (p*SNF1-I132G*), minimal YNB media with –URA amino acid supplement was applied. For the growth rate experiments cells were grown on 100 well plates in YNB with complete amino acid supplement and 4% glucose (w/v) until logarithmic phase, subcultured into fresh medium on a new 100 well plate and grown until logarithmic phase again. 10 μl of each culture was resuspended in 250 μl of fresh YNB medium with 4% or 0.2% glucose (w/v) on a new plate and cultivated in Bioscreen analyser C for 96 h at $30^0$C or $22^0$C. OD measurements at 600 nm were taken every 10 min with prior shaking. Each strain was represented in sextuplicates.

For microscopy experiments on the BY4741 wild type and/or cells with genetically integrated fluorescent proteins, minimal YNB media (1.7 g/l yeast nitrogen base without amino acids and $(NH_4)_2SO_4$, 5 g/l $(NH_4)_2SO_4$, 0.79 g/l complete amino acid supplement as indicated by manufacturer) with appropriate glucose concentrations was used. In brief, cells were first streaked onto YPD plates, grown overnight at 30ºC prior to culturing in liquid minimal YNB media with complete amino acid supplement and 4% glucose overnight, then sub-culturing into fresh YNB with 4% glucose for 4h with shaking at 30ºC. Cultures were spun at 3,000rpm, re-suspended into fresh YNB with or without glucose, immobilized in 1μl spots onto an 1% agarose well perfused with YNB minimal media with an appropriate glucose concentration enclosed between a plasma-cleaned BK7 glass microscope coverslip and slide, which permitted cells to continue to grow and divide (19,20) while being observed for up to several hours if required.

**SDS-PAGE**
50 ml cultures of YSH1703 transformed with centromeric pMig1-HA and pSNF1-I132G-TAP or pSNF1-TAP plasmids were grown until mid-log phase in yeast nitrogen base, 4% glucose, uracil and histidine deficient. Each culture was separated into two new cultures with 4% and 0.05% glucose, respectively, and incubated for 30 min. The following procedure was adapted from Bendrioua et al.(16). Cells were harvested by centrifugation (3,000rpm, 50s), suspended in 1 ml of 0.1M NaOH for 5 min and spun down. Pellets were suspended in 2 ml of 2M NaOH with 7% β- mercaptoethanol for 2 min and then 50% trichloroacetic acid was added. Samples were vortexed and spun down at 13,000rpm. The pellets were washed in 0.5 ml of 1M Tris-HCl (pH 8.0), resuspended in 50 μl of 1x SDS sample buffer (62.5 mM Tris-HCl (pH 6.8), 3% SDS, 10% glycerol, 5% β-mercaptoethanol, and 0.004% bromophenol blue) and boiled for 5 min. The protein extracts were obtained by centrifuging at the maximal speed and collecting the supernatants. For western blotting, 50 μg of extracted proteins were resolved on a Criterion TGX 10% precast polyacrylamide gel, then transferred onto a nitrocellulose membrane (Trans-Blot Turbo Transfer Pack, Bio-Rad) using Trans-Blot Turbo Transfer System (Bio-Rad). After transfer, the membrane was blocked in Odyssey Blocking buffer (LI-COR Biosciences). Mig1 was detected using primary mouse anti-HA (1:2000) antibodies (Santa Cruz), then secondary goat anti-mouse IRDye-800CW (1:5000) antibodies (LI-COR Biosciences). The result was visualized by using an infrared imager (Odyssey, LI-COR Biosciences), 800nm channel.

**Native PAGE**
A 50 ml culture of the YSH2862 strain was grown until mid-log phase in rich media with 4% glucose, then, 25 ml of the culture was transferred into fresh YPD with 4% glucose, and the rest into YPD with 0.05% glucose for 30 min. The cultures were harvested by centrifugation, suspended in 0.1ml of solubilization buffer (100 mM Tris-HCl, pH 6.8, 0.1 mM $Na_3VO_4$ , 1x protease inhibitor cocktail (Roche), 0.1% Triton-X100). 400μl of glass beads were added, and cells were broken by FastPrep, 6m/s, 20 s. Protein extracts were obtained by adding 150 μl of solubilization buffer, centrifugation at 13,000 rpm, 5min and collecting the supernatant. Protein quantification was performed by using Bradford with BSA standard (Bio-Rad). 250 μg of total protein extracts were run on a Criterion TGX Stain Free 10% precast polyacrylamide gel (Bio-Rad). Samples were diluted 1:1 with 2x Native Sample Buffer (Bio-Rad). Electrophoresis was performed at 4$^0$C starting at 100V until the bromophenol blue line reached the end of the gel. The gel was transferred onto a nitrocellulose membrane (Trans-Blot Turbo Transfer Pack, Bio-Rad) using Trans-Blot Turbo Transfer System (Bio-Rad). After transfer, the membrane was blocked in Odyssey Blocking buffer (LI-COR Biosciences), analyzed by immunoblotting with mouse anti-GFP (1:500) antibodies (Roche) and visualized with goat anti-mouse IRDye-800CW (1:5,000) antibodies (LI-COR Biosciences) by using an infrared imager (Odyssey, LI-COR Biosciences), 800nm channel. As a molecular weight reference, a NativeMark Unstained Protein Standards (Invitrogen) were used.

**Slimfield microscopy**
GFP and mCherry excitation used co-aligned linearly polarized 488 nm and 561 nm wavelength 50 mW lasers (Coherent Obis) respectively which could be attenuated independently via neutral density filters followed by propagation through an achromatic λ/2 plate to rotate the plane of polarization prior to separation into two independent paths generated by splitting into orthogonal polarization components by a polarization splitting cube to enable simultaneous Slimfield illumination and a focused laser bleach illumination path for fluorescence recovery after photobleaching (FRAP) when required. The two paths were reformed into a single common path via a second polarization cube, circularized for polarization via an achromatic λ/4 plate with fast axis orientated at 45º to the polarization axes of each path and directed at ~6 W/cm$^2$ excitation intensity onto the sample mounted on an *xyz* nanostage (Mad City Labs) via a dual-pass green/red dichroic mirror centered at long-pass wavelength 560nm and emission filters with 25nm bandwidths centered at 525nm and 594nm (Chroma).

Fluorescence emissions were captured by a 1.49NA oil immersion objective lens (Nikon) and split into green and red detection channels using a bespoke color splitter utilizing a long-pass dichroic mirror with wavelength cut-off of 565nm prior to imaging each channel onto separate halves of the same EMCCD camera detector (iXon DV860-BI, Andor Technology, UK) at a pixel magnification of 80 nm/pixel using 5ms camera exposure time. We confirmed negligible measured crosstalk between GFP and mCherry signals to red and green channels respectively, using purified GFP and mCherry sampled in an *in vitro* surface immobilization assay (details below).

**Microfluidics control of single cell imaging**
To investigate time-resolved glucose concentration-dependent changes in Mig1-GFP localization in individual yeast cells, we used bespoke microfluidics and our bespoke control software *CellBild* (LabVIEW, National Instruments), enabling cell-to-cell imaging in response to environmental glucose changes. *CellBild* controlled camera acquisition synchronized to flow-cell environmental switches via a syringe pump containing an alternate glucose environment. Microfluidic flow-chambers were based on an earlier 4-channel design (62).

Prior to each experiment flow-chambers were wetted and pre-treated for 15min with 1 mg/ml of concanavalin A (ConA) which binds to the glass surface of the plasma cleaned flow-chamber. Cells were introduced via a side channel and were left bind to ConA for 15min to immobilize cells on the surface. Any remaining ConA and unbound cells were washed out and a steady flow of YNB with 0% glucose provided to one of the central channels by gravity feed. A syringe pump synchronized with image acquisition introduced YNB with 4% glucose in the second central channel. The pumped alternate environment reaches cells within 1-2s at a flow rate of 10 μl/min, enabling rapid change between two different glucose concentrations.

Slimfield imaging was performed on a similar bespoke microscope setup at comparable laser excitation intensities and spectral filtering prior to imaging onto a Photometrics *Evolve Delta 512* EMCCD camera at 200 frames per second. Alternating frame laser excitation (ALEX) was used to minimize any autofluorescence contamination in the red channel introduced by the blue excitation light.

**Foci detection, tracking and stoichiometry determination**
Our bespoke foci detection and tracking software objectively identifies candidate bright foci by a combination of pixel intensity thresholding and image transformation to yield bright pixel coordinates. The intensity centroid and characteristic intensity, defined as the sum of the pixel intensities inside a 5 pixel radius region of interest around the foci minus the local background and corrected for non-uniformity in the excitation field are determined by iterative Gaussian masking. If the signal-to-noise ratio of the foci, defined as the characteristic intensity per pixel/background standard deviation per pixel, is >0.4 it is accepted and fitted with a 2D radial Gaussian function to determine its sigma width, which our simulations indicate single-molecule sensitivity under typical *in vivo* imaging conditions (26). Foci in consecutive image frames within a single point spread function (PSF) width, and not different in brightness or sigma width by more than a factor of two, are linked into the same track. The microscopic diffusion coefficient $D$ is then estimated for each accepted foci track using mean square displacement analysis, in addition to several other mobility parameters.

Cell and nuclear boundaries were segmented from GFP and mCherry fluorescence images respectively using a relative threshold pixel intensity value trained on simulated images of uniform fluorescence in idealized spherical compartments. An optimized threshold value of 0.3 times the mean compartment fluorescence intensity segmented the boundary to within 0.5 pixels.

The characteristic brightness of a single GFP molecule was determined directly from *in vivo* data and corroborated using *in vitro* immobilized protein assays (21). The intensity of tracked fluorescent foci in live cells was measured over time as described above. These followed an approximately exponential photobleach decay function of intensity with respect to time. Every oligomeric Mig1-GFP complex as it photobleaches to zero intensity will emit the characteristic single GFP intensity value, $I_{GFP}$, i.e. the brightness of a single GFP molecule, given in our case by the modal value of all foci intensities over time, and can potentially bleach in integer steps of this value at each sampling time point. This value of $I_{GFP}$ was further verified by Fourier spectral analysis of the pairwise distance distribution (21) of all foci intensities which yields the same value to within measurement error in our system.

All foci tracks found within 70 image frames of the start of laser illumination were included in the analysis and were corrected for photobleaching by weighting the measured foci intensity $I$ at a time $t$ following the start of laser illumination with a function $\exp(+t/t_b)$ to correct for the exponential photobleach decay $I_0\exp(-t/t_b)$, of each intensity trace with a fixed time constant, where $I_0$ is the initial unbleached intensity. This photobleach time constant $t_b$ was determined from exponential decay fits to the foci intensities and whole cell intensities over time to be 40 ± 0.6 ms. Stoichiometries were obtained by dividing the photobleach estimate for the initial intensity $I_0$ of a given foci by the characteristic single GFP molecule brightness value $I_{GFP}$.

Autofluorescence correction was applied to pool quantification by subtracting the red channel image from the green channel image multiplied by a correlation factor. By comparing wild type and GFP cell images we confirmed that when only the GFP exciting 488 nm wavelength laser was used the green channel image contained fluorescence intensity from GFP and autofluorescence, while the red channel contains only autofluorescence pixels, consistent with expectations from transmission spectra of known autofluorescent components in yeast cells. We measured the red channel autofluorescence pixels to be linearly proportional to the green channel autofluorescence pixels. The scaling factor between channels was determined by Slimfield imaging of the wild type yeast strain (i.e. non GFP) under the same conditions and comparing intensity values pixel-by-pixel in each channel. A linear relationship between pixels was found with scaling factor of $0.9 \pm 0.1$.

Copy numbers of Mig1-GFP of the pool component were estimated using a previously developed CoPro algorithm (26). In brief, the cytoplasmic and nuclear pools were modelled as uniform fluorescence over spherical cells and nuclei using experimentally measured radii. A model PSF was integrated over these two volumes to create model nuclear and cytoplasmic images and then their relative contributions to the camera background and autofluorescence corrected GFP intensity image determined by solving a set of linear equations for each pixel. Dividing the contributions by the characteristic single GFP molecule intensity and correcting for out-of-plane foci yields the pool concentration.

Stoichiometry distributions were rendered as objective kernel density estimations (21) using a Gaussian kernel with bandwidth optimized for normally distributed data using standard MATLAB routines.

**Stochastic optical reconstruction microscopy (STORM)**
To photoswitch Mig1-mEos2, a 405 nm wavelength laser (Coherent Obis), attenuated to ~1mW/ cm$^2$ was used in conjunction with the 488 nm and 561 nm lasers on the Slimfield microscope, similar to previous super-resolution imaging of yeast cells (63). The 405 nm laser light causes mEos2 to photoswitch from a green (excitable via the 488 nm laser) to a red (excitable by the 561 nm laser) fluorescent state. Using low intensity 405 nm light generates photoactive fluorophore foci, photobleached by the 561 nm laser at a rate which results in an approximately steady-state concentration density in each live cell studied. The bright foci were tracked similar to the Slimfield data but were used to generate a super-resolved image by the summation of 2D Gaussian functions at each tracked Mig1-mEos2 localization in time with a width of ~40 nm, the measured lateral precision following automated particle tracking (26).

**Fluorescent protein brightness characterization**
We used a surface-immobilization assay described previously (20,26) employing antibody conjugation to immobilize single molecules of GFP respectively onto the surface of plasma-cleaned BK7 glass microscope coverslips and imaged using the same buffer medium and imaging conditions as for live cell Slimfield experiments, resulting in integrated single-molecule peak intensity values for mGFP of $4,600 \pm 3,000$ ($\pm$ half width half maximum, HWHM) counts. Similar experiments on unmodified purified Clontech eGFP generated peak intensity values of $4,700 \pm 2,000$ counts, statistically identical to that of mGFP (Student $t$-test, $p < 0.2$) with no significant indication of multimerization effects from the measured distribution of foci intensity values. Similarly, Slimfield imaging and foci stoichiometry analysis on Mig1-mGFP and Mig1-eGFP cell strains were compared *in vivo* under high and low glucose conditions in two separate cell strains, resulting in distributions which were statistically identical (Pearson's $\chi^2$ test, p<0.001). These results indicated no measurable differences between multimerization state or single-molecule foci intensity between mGFP and eGFP which enabled direct comparison between Mig1-eGFP cell strain data obtained from preliminary experiments here and from previous studies (16).

Maturation effects of mCherry and GFP were investigated by adding mRNA translation inhibitor antibiotic cycloheximide, final concentration 100 µg/ml, for 1h (64), photobleaching cells, then monitoring any recovery in fluorescence as a metric for newly matured fluorescent material in the cell. Cells were prepared for microscopy as before but using cycloheximide in all subsequent preparation and imaging media and imaged using a commercial mercury-arc excitation fluorescence microscope Zeiss Axiovert 200M (Carl Zeiss MicroImaging) onto an ApoTome camera using a lower excitation intensity than for Slimfield imaging but a larger field of view, enabling a greater number of cells to be imaged simultaneously.

Surface-immobilized cells using strain YSH2863 were photobleached by continuous illumination for between 3min 40s to 4min until dark using separate filter sets 38HE and 43HE for GFP and mCherry excitation, respectively. Fluorescence images were acquired at subsequent time intervals up to 120min and analyzed using AxioVision software (fig. S6). The background-corrected total cellular fluorescence intensity was quantified at each time point for each cell using ImageJ software. Comparison between Mig1-GFP fluorescence signal and the green channel signal from the parental strain BY4741, and the Nrd1-mCherry signal and the red channel signal from the parental strain, indicate fluorescence recovery after correction above the level of any autofluorescence contributions of <15% for GFP and mCherry over the timescale of our experiments, consistent with previous estimates of *in vivo* maturation times for GFP and mCherry (20,21,65).

**Characterizing Mig1-GFP clusters** *in vitro*
Using Slimfield microscopy under the same imaging conditions as for live cell microscopy we measured the fluorescent foci intensity of 1μg/ml solutions of purified Mig1-mGFP and mGFP using the normal imaging buffer of PBS, compared with the imaging buffer supplemented with 1kDa molecular weight PEG at a concentration of 10% (w/v) used to reproduce cellular depletion forces (34,60).

**Circular dichroism measurements**
Purified Mig1-mGFP was placed in 25 mM $Na_2HPO_4$, pH 7.0, by buffer exchange procedure with a Pur-A-Lyser Maxi dialysis Kit (Sigma Aldrich) for 3h at $4^0$C with constant stirring in 500 ml a buffer. Circular dichroism measurements were performed on a Jasco J810 circular dichromator with Peltier temperature control and Biologic SFM300 stop-flow accessory on 0.16mg/ml Mig1-mGFP samples with or without 20% PEG-1000 at $20^0$C, from 260 to 200 nm, a 2 nm band width, 2 sec response time, at the speed of 100 nm/min. The resulting spectrum represents the average of 5 scans, indicating a typical SD error of ~0.1 mdeg ellipticity. Spectra from 25 mM $Na_2HPO_4$ and 25 mM $Na_2HPO_4$ with 20% (w/v) PEG were used as a background and subtracted from spectra of Mig1-mGFP without or with 20% (w/v) PEG respectively.

**Immuno-gold electron microscopy**
Cells for Mig1-GFP and Msn2-GFP strains as well as the wild type control strain containing no GFP were grown using the same conditions as for Slimfield imaging but pelleted down at the end of growth and prepared for immuno electron microscopy using an adaptation of the Tokuyasu cryosectioning method (66) following the same protocol that had been previously optimized for budding yeast cells (67) to generate ~90nm thick cryosections, with the exception that the sections were picked up on a drop of 2.3M sucrose, placed on the grid, then floated down on PBS, and then immunolabeled immediately, rather than storing on gelatine as occurred in the earlier protocol. The grids used were nickel, with a formvar/carbon support film. In brief, the immunolabeling protocol used a 0.05M glycine in PBS wash of each section for 5 min followed by a block of 10% goat serum in PBS (GS/PBS) pre-filtered through a 0.2 μm diameter filter. Then an incubation of 1 h with the primary antibody of rabbit polyclonal anti-GFP (ab6556, Abcam) at 1 in 250 dilution from stock in GS/PBS. Then five 3 min washes in GS/PBS. Then incubation for 45 min with the goat anti-IgG-rabbit secondary antibody labeled with 10nm diameter gold (EM.GAR10, BBI solutions) at a dilution of 1 in 10 from stock. Sections were then washed five more times in GS/PBS prior to chemical fixation in 1% glutaraldehyde in sodium phosphate buffer for 10 min, then washed in dH$_2$0 five times for 3 min each and negative-stained using methyl cellulose 2% in 0.4% uranyl acetate, and then washed twice more in dH$_2$0 prior to drying for 10 min. Drop sizes for staining, blocking and washing onto sections were 50 μl, while antibody incubations used 25 μl drops, all steps performed at room temperatures.

Electron microscopy was performed on these dried sections using a 120kV Tecnai 12 BioTWIN (FEI) electron microscope in transmission mode, and imaged onto an SIS Megaview III camera. Control cells containing no GFP showed no obvious signs of gold labeling. Mig1-GFP and Msn2-GFP strains showed evidence for gold labeling, though the overall labeling efficiency was relatively low with several images containing only a single gold foci per cell with labeling largely absent from the nucleus possibly due to poor antibody accessibly into regions of tightly packed DNA. However, we observed 10 cells from a set of ~150 from each of the Mig-GFP and Msn2-GFP strains (i.e. ~7% of the total) which showed >1 gold foci clustering together inside an area of effective diameter ~50nm or less, with up to 7 gold foci per cluster being observed.

**Bioinformatics analysis and structural modeling**

Bioinformatics analysis was used to identity candidate promoter sequences in the budding yeast genome. The Mig1 target pattern sequence was identified based on 14 promoter sequences (35) using the IUPAC nucleotide code. The entire *S. cerevisiae* S288c genome was scanned in order to find all sequences that matched the pattern. The scanning was performed by RNABOB software (68), and collated for any further analysis and identification of the sequences lying within promoter regions. All information regarding *S. cerevisiae* genes was obtained from SGD database (http://yeastgenome.org/).

We used bioinformatics to investigate the extent of intrinsic disorder in the amino acid sequence of budding yeast Mig1 and Msn2 proteins as well as the *E. coli lac* repressor LacI, employing the Predictor of Natural Disordered Regions (PONDR) algorithm (69) (online tool http://www.pondr.com/cgi-bin/PONDR/pondr.cgi) with a VL-XT algorithm. We also used the secondary structure prediction algorithm of PyMOL (http://www.pymolwiki.org/index.php/Dss) to highlight disordered and structured regions and display the unfolded protein chain, and used PSI-BLAST multiple sequence alignment to determine conserved structural features of Mig1 for the zinc finger motif in combination with the DISOPRED (46) algorithm as a comparison to PONDR, which produced very similar results (online tool http://www.yeastrc.org/pdr/).

**Oligomerization state of Mig1-GFP in the 'pool'**

Experimental *in vitro* assays of surface immobilized GFP coupled to simulations trained on these single-molecule intensity measurements but using noise levels comparable to *in vivo* cellular imaging conditions (26) indicate single-molecule sensitivity of GFP detection under our millisecond imaging conditions. However, if the nearest neighbor separation of individual GFP 'foci' are less than the optical resolution limit $w$ of our microscope (which we measure as ~230 nm for GFP imaging) then distinct fluorescent foci will not be detected and instead will be manifest as a diffusive 'pool'.

If each GFP 'foci' in the pool has a mean stoichiometry $S$ then the mean number of GFP foci, $F$, in the pool is $n_{pool}/S$ and the 'pool' condition for nearest neighbor foci separation $s$ indicates that $s<w$.

The estimated range of mean total pool copy number from nucleus and cytoplasm combined, $n_{pool}$, is ~590-1,100 molecules depending on extracellular glucose conditions.

Approximating the cell volume as equal to the combined volumes of all uniformly separated foci in the pool (equal to the total number of foci multiplied by the volume of an equivalent sphere of radius $r$) indicates that $F.4\pi r^3/3 = 4\pi d^3/3$, thus $r = d/F^{1/3}$, where we use the mean measured cell diameter $d$ of ~5 μm.

However, mobile foci with a microscopic diffusion coefficient $D$ will diffuse a mean two-dimensional distance $b$ in focal plane of $(4D.\Delta t)^{1/2}$ in a camera sampling time window $\Delta t$ of 5 ms. Using $D$ ~6 μm$^2$ s$^{-1}$ as a lower limit based on the measured diffusion of low stoichiometry cytoplasmic Mig1-GFP foci detected indicates $b$ ~340 nm so the movement-corrected estimate for $s$ is $r-b$, thus $s < w$ indicates that $r < b+w$, or $d/F^{1/3} < b+w$.

Therefore, $d(S/n_{pool})^{1/3} < b+w$, and $S < n_{pool}((b+w)/d)^3$. Using ~590-1,100 molecules from the measured mean range of $n_{pool}$ indicates that the upper limit for $S$ is in the range 0.8-1.4; in other words, Mig1-GFP foci in the pool are consistent with being a monomer.

**Analysis of the mobility of foci**

For each accepted foci track the mean square displacement (MSD) was calculated from the optimized intensity centroid at time $t$ of $(x(t),y(t))$ assuming a tracks of $N$ consecutive image frames at a time interval $\tau = n\Delta t$ is (70,71) where $n$ is a positive integer is:

$$MSD(\tau) = MSD(n\Delta t) = \frac{1}{N-1-n} \sum_{i=1}^{N-1-n} \{[x(i\Delta t + n\Delta t) - x(i\Delta t)]^2 + [y(i\Delta t + n\Delta t) - y(i\Delta t)]^2\}$$
$$= 4D\tau + 4\sigma^2$$

Here $\sigma$ is the lateral ($xy$) localization precision which we estimate as ~40 nm (26). The microscopic diffusion coefficient $D$ was then estimated from the gradient of a linear fit to the first four time interval data points of the MSD vs $\tau$ relation for each accepted foci track.

To determine the proportion of mobile and immobile Mig1-GFP fluorescent foci we adapted an approach based on cumulative probability-distance distribution analysis (12). Here we generated cumulative distribution functions (CDFs) for all nuclear and cytoplasmic tracks, such that the CDF in each dataset is the probability distribution function $p_c$ associated with $r^2$, the square of the displacement between the first and second data points in each single track, which was generated for each track by calculating the proportion of all tracks in a dataset which have a value of $r^2$ less than that measured for that one track. The simplest CDF model assumes a Brownian diffusion propagator function $f(r^2)$ for a single effective diffusion coefficient component of:

$$f(r^2) = \frac{1}{4\pi D \Delta t} \exp\left(\frac{r^2}{4D\Delta t}\right)$$

Here, $D$ is the effective diffusion coefficient and $\Delta t$ is image sampling time per frame (i.e. in our case 5 ms). This gives a CDF single component solution of the form:

$$p_c(r^2) = 1 - \exp\left(\frac{r^2}{4D\Delta t}\right)$$

We investigated both single and more complex multi-component CDF models using either 1,2 or 3 different $D$ values in a weighted sum model of:

$$p_c(r^2) = \sum_{i=1}^{n} A_i \left(1 - \exp\left(\frac{r^2}{4D_i \Delta t}\right)\right)$$

Here $n$ is 1, 2 or 3. Multi-component fits were only chosen if they lowered the reduced $\chi^2$ by >10%. For cytoplasmic foci at *glucose* (+/-) and for nuclear foci at *glucose* (-) this indicated single component fits for diffusion coefficient with a $D$ of ~1-2 μm$^2$/s, whereas nuclear foci at *glucose* (+) were fitted using two components of $D$, ~20% with a relatively immobile component, $D$ ~0.1-0.2 μm$^2$/s, and the remainder a relatively mobile component, $D$ ~1-2 μm$^2$/s, while using three components produced no statistically significant improvement to the fits. These values of $D$ agreed to within experimental error to those obtained using a different method which fitted two analytical Gamma functions to the distribution of all calculated microscopic diffusion coefficients of tracked foci in the nucleus at *glucose* (+), which assumed a total probability distribution function $p_\gamma$ of the form: (28)

$$p_\gamma(x, D) = \sum_{i=1}^{2} \frac{A_i (m/D)^m x^{n-1} \exp(-mx/D)}{(m-1)!}$$

Here, $m$ is the number of steps in the MSD *vs* $\tau$ trace for each foci track used to calculate $D$ (i.e. in our instance $m=4$).

We also probed longer time scale effects on foci mobility for each accepted foci trajectory. Here, average MSD values were generated by calculating mean MSD values for corresponding time interval values across all foci trajectories in each dataset, but pooling traces into low stoichiometry (≤ 20 Mig1-GFP molecules per foci) and high stoichiometry (> 20 Mig1-GFP molecules per foci). We compared different diffusion models over a 30 ms time interval scale, corresponding to the shortest time interval range from any of the mean MSD trace datasets.

We found in all cases that mean MSD traces could be fitted well ($\chi^2$ values in the range 1-12) using a subdiffusion model of precision-corrected MSD = $4\sigma^2 + 4D\tau^\alpha$, where $\alpha$ the anomalous diffusion coefficient and $D$ is the microscopic lateral diffusion coefficient. Optimized fits indicated values of $D$ in the range 0.08-0.2 μm$^2$/s and those for $\alpha$ of ~0.4-0.8. Corresponding fits to a purely Brownian diffusion model (i.e. $\alpha = 1$) generated much poorer fits ($\chi^2$ values in the range 4-90).

**Analyzing trans-nuclear tracks**
The segmentation boundary output for the nucleus was fitted with a smoothing spline function, with smoothing parameter $p = 0.9992$ to sub-pixel precision. Trajectories which contained points on either side of the nuclear boundary were considered trans-nuclear. The crossing point on the nuclear boundary was found by linearly interpolating between the first pair of points either side of the nuclear boundary. Coordinates were normalized to this point and the crossing time and were rotated such that $y'$ and $x'$ lie perpendicular and parallel to the membrane crossing point.

**Investigating Mig1-GFP molecular turnover**
Turnover of Mig1-GFP was investigated using fluorescence recovery after photobleaching (FRAP). In brief a 200 ms 10m W focused laser beam pulse of lateral width ~1 μm was used to photobleach the fluorescently-labelled nuclear contents on a cell-by-cell basis and then ≤ 10 Slimfield images were recorded over different timescales spanning a range from 100 ms to ~1,000 s. The copy number of pool and foci in each image at subsequent time points $t$ post focused laser bleach was determined as described and corrected for photobleaching. These post-bleach photoactive Mig1-GFP copy number values $C(t)$ could then be fitted using a single exponential recovery function:

$$C(t) = C(0)(1 - \exp(-t/t_R))$$

Where $t_R$ is the characteristic recovery (i.e. turnover) time (19). These indicated a value of 133 ± 20 s (±SEM) for nuclear foci at glucose (+), and 3 ± 14 s for nuclear pool at *glucose* (+) and (-).

**Modeling the effective diameter of clusters**
The effective diameter $d$ of a cluster was estimated from the measured point spread function width $pf_{foci}$ (defined at twice sigma value of the equivalent Gaussian fit from our single particle tracking algorithm) corrected for the blur due to particle diffusion in the camera exposure time of $\Delta t$ as:

$$d = p_{foci} - p_{GFP} - \sqrt{4D\Delta t}$$

Where $D$ is the measured microscopic diffusion coefficient for that track and $p_{GFP}$ is the measured point spread function width of surface-immobilized GFP (i.e. twice the sigma width of 230nm measured in our microscope, or 460nm). We explored a heuristic packing model of $d \sim S^a$ for Mig1-GFP monomers in each cluster, such that a tightly packed spherical cluster of volume $V$ composed of $S$ smaller ca. spherical monomers each of volume $V_1$ and diameter $d_1$ varied as $V = S.V_1$ thus $4\pi(d/2)^3 = S.4\pi(d_1/2)^3$, thus in the specific instance of a perfect spherical cluster model $a = 1/3$.

In principle, for general shapes of clusters for different packing conformations we expect $0 \leq a \leq 1$ such that e.g. if clusters pack as a long, thin rod of Mig1 monomers which rotates isotropically during time $\Delta t$, then $a = 1$. Whereas, if Mig1 monomers bind to a putative additional 'anchor' type structure to occupy available binding sites in forming a cluster, such that the size of the cluster does not significantly change with $S$ but is dependent on the size of the putative anchor structure itself, then $a = 0$. Our optimized fits indicate $a = 0.32 \pm 0.06$ (±SEM), i.e. consistent with an approximate spherical shape cluster model.

**Modeling the probability of overlap in *in vitro* fluorescent protein characterization**
The probability that two or more fluorescent protein foci are within the diffraction limit of our microscope in the *in vitro* characterization assays was determined using a previously reported Poisson model (25) to be ~10% at the *in vitro* protein concentrations used here. Such overlapping fluorescent proteins are detected as higher apparent stoichiometry foci.

**Software and DNA sequence access**
All our bespoke software developed, and Mig1 secondary structure prediction 3D coordinates pymolMig1.pdb, are freely and openly accessible via https://sourceforge.net/projects/york-biophysics/. The bespoke plasmid sequence information for the GFP reporter is openly accessible via https://www.addgene.org/75360/.

**Statistical tests**
All statistical tests used are two-sided unless stated otherwise.

| Strain name | Background | Genotype | Source/Reference |
|---|---|---|---|
| YSH1351 | S288C | *MATa HIS3D0 LEU2D1 MET15D0 URA3D0* | S. Hohmann collection |
| YSH1703 | W303-1A | *MATa mig1Δ::LEU2 snf1Δ::KanMX* | S. Hohmann collection |
| YSH2267 | BY4741 | *MATa his3D1 leu2D0 met15D0 ura3D0 mig1Δ::KanMX NRD1-mCherry-hphNT1* | S. Hohmann collection |
| YSH2350 | BY4741 | *MATa MSN2-GFP-HIS3 NRD1-mCherry-hphNT1 MET LYS* | (48) |
| YSH2856 | BY4741 | *MATa MIG1-eGFP-KanMX NRD1-mCherry-HphNT1 snf1Δ::LEU2 MET LYS* | This study |
| YSH2348 | BY4741 | *MATa MIG1-GFP-HIS3 NRD1-mCherry-hphNT1 MET LYS* | (16) |
| YSH2862 | BY4741 | *MATa MIG1-GFPmut3-HIS3* | This study |
| YSH2863 | BY4741 | *MATa MIG1-GFPmut3-HIS3 NRD1-mCherry-HphMX4* | This study |
| YSH2896 | BY4741 | *MATa MIG1-mEOs2-HIS3* | This study |
| ME404 | BY4741 | *"BY4741 MSN2-mKO2::LEU2 MIG1-mCherry::spHIS5 GSY1-24xPP7::KANMX msn4Δ mig2Δ nrg1::HPHMX nrg2::Met15 SUC2::NatMX"* | (17) |
| ME412 | BY4741 | *BY4741 MSN2-mKO2::LEU2 MIG1(Δaa36-91)-mCherry::spHIS5 GSY1-24xPP7::KANMX msn4Δ mig2Δ nrg1::HPHMX nrg2::Met15* | (17) |

| Plasmid name | Description | Source/Reference |
|---|---|---|
| pMIG1-HA | *HIS3* | (72) |
| pSNF1-TAP | *URA3, in pRS316* | S. Hohmann collection |
| pSNF1-I132G-TAP | *URA3, in pRS316* | S. Hohmann collection |
| pmGFPS | *HIS3, GFPmut3 S65G, S72A, A206K* | This study |
| pMig1-mGFP | *6xHIS-Mig1-GFPmut3 in pRSET A* | This study |
| pmEOs2 | *mEOs2-HIS3 in pMK-RQ* | This study |
| YDp-L | *LEU2* | (73) |
| YDp-H | *HIS3* | (73) |
| BM3726 | *Mig1 (Ser222,278,311,381 → Ala), URA3, in pRS316* | M. Johnston collection (27) |
| pDZ276 | *PP7-2xGFP::URA3* | (17) |

**Table S1.** *S. cerevisiae* **cell strains and plasmids**. List of all strains and plasmids used in this study.

| Cell Strain | Gluc. (+/-) | | Cyt. Pool | Nuc. Pool | Tot. Pool | Cyt. Foci | Nuc. Foci | Tot. Foci | Tot. Cyt. | Tot. Nuc. |
|---|---|---|---|---|---|---|---|---|---|---|
| **Mig1-GFP** | (+) | Mean | 509 | 77 | 586 | 57 | 190 | 246 | 580 | 226 |
| | | SD | 274 | 101 | 336 | 79 | 99 | 100 | 276 | 155 |
| | (-) | Mean | 949 | 140 | 1088 | 311 | 35 | 345 | 1156 | 176 |
| | | SD | 394 | 97 | 392 | 212 | 63 | 203 | 399 | 124 |
| **Mig1-GFP snf1Δ** | (+) | Mean | 947 | 807 | 1754 | 118 | 162 | 280 | 1065 | 969 |
| | | SD | 728 | 398 | 1127 | 169 | 69 | 238 | 897 | 467 |
| | (-) | Mean | 608 | 611 | 1219 | 334 | 164 | 498 | 941 | 775 |
| | | SD | 450 | 325 | 775 | 374 | 71 | 445 | 824 | 396 |
| **Msn2-GFP** | (+) | Mean | 1422 | 551 | 1973 | 333 | 81 | 414 | 1755 | 632 |
| | | SD | 977 | 608 | 1585 | 196 | 138 | 334 | 1173 | 746 |
| | (-) | Mean | 2487 | 1692 | 4179 | 776 | 320 | 1096 | 3263 | 2012 |
| | | SD | 1360 | 1221 | 2581 | 635 | 269 | 904 | 1995 | 1490 |

| Cell Strain | Gluc. (+/-) | | S, Nuc. Foci (molecules) | D, Nuc. Foci ($\mu m^2/s$) | S, Tran-nuc. (molecules) | D, Trans-nuc. Foci ($\mu m^2/s$) | S, Cyt. Foci (molecules) | D, Cyt. Foci ($\mu m^2/s$) |
|---|---|---|---|---|---|---|---|---|
| **Mig1-GFP** | (+) | Mean | 19.0 | 0.8 | 10.6 | 1.3 | 6.6 | 1.4 |
| | | SD | 16.2 | 0.8 | 10.2 | 1.2 | 4.9 | 1.4 |
| | | N | 7.2 | 7.2 | 1.0 | 1.0 | 1.1 | 1.1 |
| | (-) | Mean | 8.5 | 1.3 | 8.7 | 1.5 | 7.2 | 1.2 |
| | | SD | 4.8 | 1.5 | 5.3 | 1.6 | 3.7 | 1.2 |
| | | N | 5.8 | 5.8 | 5.1 | 5.1 | 17.8 | 17.8 |
| **Mig1-GFP snf1Δ** | (+) | Mean | 17.5 | 1.1 | 8.9 | 1.9 | 6.2 | 1.3 |
| | | SD | 10.9 | 1.1 | 6.0 | 2.0 | 2.2 | 1.2 |
| | | N | 13.2 | 13.2 | 1.2 | 1.2 | 5.0 | 5.0 |
| | (-) | Mean | 23.5 | 0.7 | 12.7 | 1.1 | 8.3 | 1.0 |
| | | SD | 15.4 | 0.8 | 6.1 | 1.4 | 4.1 | 1.2 |
| | | N | 10.9 | 10.9 | 0.5 | 0.5 | 9.1 | 9.1 |
| **Msn2-GFP** | (+) | Mean | 34.5 | 0.7 | 21.8 | 1.5 | 25.7 | 1.2 |
| | | SD | 26.6 | 0.9 | 16.7 | 1.2 | 19.5 | 1.1 |
| | | N | 3.5 | 3.5 | 1.9 | 1.9 | 4.8 | 4.8 |
| | (-) | Mean | 46.5 | 0.9 | 43.9 | 1.1 | 30.1 | 1.0 |
| | | SD | 31.6 | 0.9 | 35.0 | 1.1 | 17.5 | 1.4 |
| | | N | 4.7 | 4.7 | 0.9 | 0.9 | 4.0 | 4.0 |

**Table S2. Foci tracking data.** Upper panel: Mean average and SD of copy number in pool and foci in each compartment. Lower panel: Mean average, SD and mean number detected per cell (N) of stoichiometry values (molecules), and microscopic diffusion coefficients $D$ in each compartment.

| Chromosome | length (bp) | N sites identified | N promoter sites |
|:---:|:---:|:---:|:---:|
| I | 230218 | 41 | 1 |
| II | 813184 | 134 | 10 |
| III | 316620 | 52 | 2 |
| IV | 1531933 | 240 | 14 |
| V | 576874 | 109 | 8 |
| VI | 270161 | 58 | 4 |
| VII | 1090940 | 168 | 13 |
| VIII | 562643 | 92 | 2 |
| IX | 439888 | 94 | 8 |
| X | 745751 | 125 | 6 |
| XI | 666816 | 117 | 6 |
| XII | 1078177 | 194 | 12 |
| XIII | 924431 | 157 | 6 |
| XIV | 784333 | 135 | 3 |
| XV | 1091291 | 185 | 11 |
| XVI | 948066 | 163 | 6 |

**Table S3. Number of potential Mig1 target promoter sites per chromosome**. List of *S.cerevisiae* chromosomes indicating the length of a chromosome, total number of potential Mig1 target sites identified and then the number of sites on promoters assuming a promoter region up to 500bp upstream of a gene.

**Msn2:**

| | |
|---|---|
| Predicted residues: 704 | Number Disordered Regions: 12 |
| Number residues disordered: 394 | Longest Disordered Region: 145 |
| Overall percent disordered: 55.97 | Average Prediction Score: 0.5577 |
| Predicted disorder segment [1]-[2] | Average Strength= 0.8759 |
| Predicted disorder segment [16]-[33] | Average Strength= 0.6958 |
| Predicted disorder segment [55]-[199] | Average Strength= 0.8311 |
| Predicted disorder segment [222]-[249] | Average Strength= 0.8237 |
| Predicted disorder segment [322]-[365] | Average Strength= 0.8820 |
| Predicted disorder segment [410]-[428] | Average Strength= 0.7475 |
| Predicted disorder segment [469]-[480] | Average Strength= 0.6545 |
| Predicted disorder segment [510]-[549] | Average Strength= 0.8040 |
| Predicted disorder segment [572]-[641] | Average Strength= 0.9319 |
| Predicted disorder segment [660]-[667] | Average Strength= 0.6829 |
| Predicted disorder segment [694]-[695] | Average Strength= 0.5325 |
| Predicted disorder segment [699]-[704] | Average Strength= 0.6783 |

**Mig1:**

| | |
|---|---|
| Predicted residues: 504 | Number Disordered Regions: 9 |
| Number residues disordered: 372 | Longest Disordered Region: 95 |
| Overall percent disordered: 73.81 | Average Prediction Score: 0.7008 |
| Predicted disorder segment [1]-[12] | Average Strength= 0.8252 |
| Predicted disorder segment [25]-[33] | Average Strength= 0.6502 |
| Predicted disorder segment [77]-[171] | Average Strength= 0.8758 |
| Predicted disorder segment [173]-[240] | Average Strength= 0.9051 |
| Predicted disorder segment [242]-[249] | Average Strength= 0.5554 |
| Predicted disorder segment [254]-[272] | Average Strength= 0.7890 |
| Predicted disorder segment [292]-[310] | Average Strength= 0.8225 |
| Predicted disorder segment [327]-[386] | Average Strength= 0.8355 |
| Predicted disorder segment [423]-[504] | Average Strength= 0.9136 |

**LacI:**

| | |
|---|---|
| Predicted residues: 360 | Number Disordered Regions: 8 |
| Number residues disordered: 149 | Longest Disordered Region: 48 |
| Overall percent disordered: 41.39 | Average Prediction Score: 0.4418 |
| Predicted disorder segment [1]-[4] | Average Strength= 0.6245 |
| Predicted disorder segment [18]-[52] | Average Strength= 0.6710 |
| Predicted disorder segment [55]-[81] | Average Strength= 0.7443 |
| Predicted disorder segment [88]-[100] | Average Strength= 0.5841 |
| Predicted disorder segment [186]-[187] | Average Strength= 0.5429 |
| Predicted disorder segment [238]-[256] | Average Strength= 0.6208 |
| Predicted disorder segment [258]-[258] | Average Strength= 0.5028 |
| Predicted disorder segment [313]-[360] | Average Strength= 0.8331 |

**Phosphorylation sites of Mig1 and Msn2 (uniprot.org, accessed February, 2016):**

| Mig1 Phosphorylation site | Disorder segment | Msn2 Phosphorylation site | Disorder segment |
|---|---|---|---|
| S264 | [254]-[272] | S194 | [55]-[199] |
| S278 | - | S201 | - |
| T280 | - | S288 | - |
| S302 | [292]-[310] | S304 | - |
| S311 | [292]-[310] | S306 | - |
| S314 | - | S308 | - |
| S80 | [77]-[171] | S432 | - |
| S108 | [77]-[171] | S451 | - |
| S214 | [173]-[240] | S582 | [572]-[641] |
| S218 | [173]-[240] | S620 | [572]-[641] |
| S222 | [173]-[240] | S625 | [572]-[641]] |
| S303 | [292]-[310] | T627 | [572]-[641] |
| S310 | [292]-[310] | S629 | [572]-[641] |
| S350 | [327]-[386] | S633 | [572]-[641] |
| S367 | [327]-[386] | | |
| S370 | [327]-[386] | | |
| T371 | [327]-[386] | | |
| S377 | [327]-[386] | | |
| S379 | [327]-[386] | | |
| S381 | [327]-[386] | | |
| S400 | - | | |
| S402 | - | | |
| T455 | [423]-[504] | | |

**Table S4. Bioinformatics analysis for intrinsically disordered sequences.** Predictions for the presence of intrinsically disordered sequences in Mig1, Msn2 and LacI, and of the positions of phosphorylation sites in Mig1 and Msn2.

**Supplementary Movie Stills and Captions**

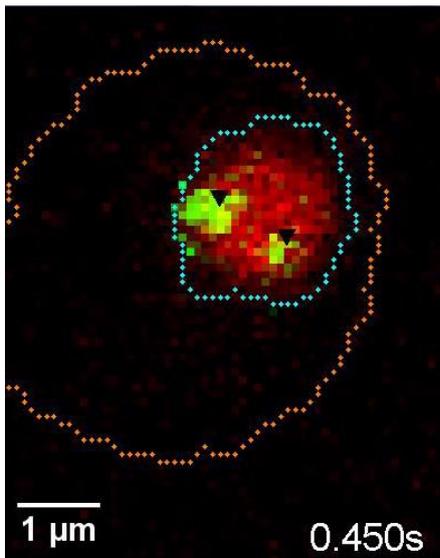

**Movie S1. Dual-color fluorescence microscopy assay at *glucose* (+).** Example cell showing *glucose* (+) nuclear Mig1-GFP localization (green, distinct foci black arrows), Nrd1-mCherry (red) and segmented cell body (orange) and nuclear envelope (cyan) indicated, slowed 15x.

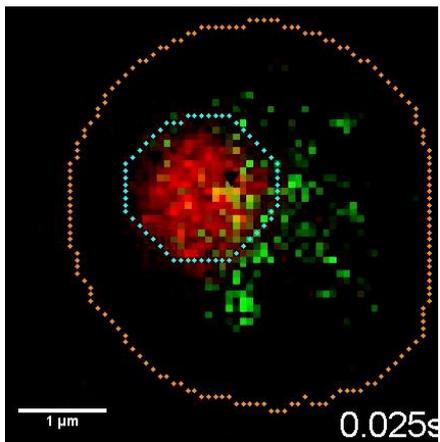

**Movie S2. Dual-color fluorescence microscopy assay at *glucose* (-).** Example cell showing *glucose* (-) Mig1-GFP localization (green, distinct foci black arrows), Nrd1-mCherry (red) and segmented cell body (orange) and nuclear envelope (cyan) indicated, slowed 200x.

# Supplementary Figures

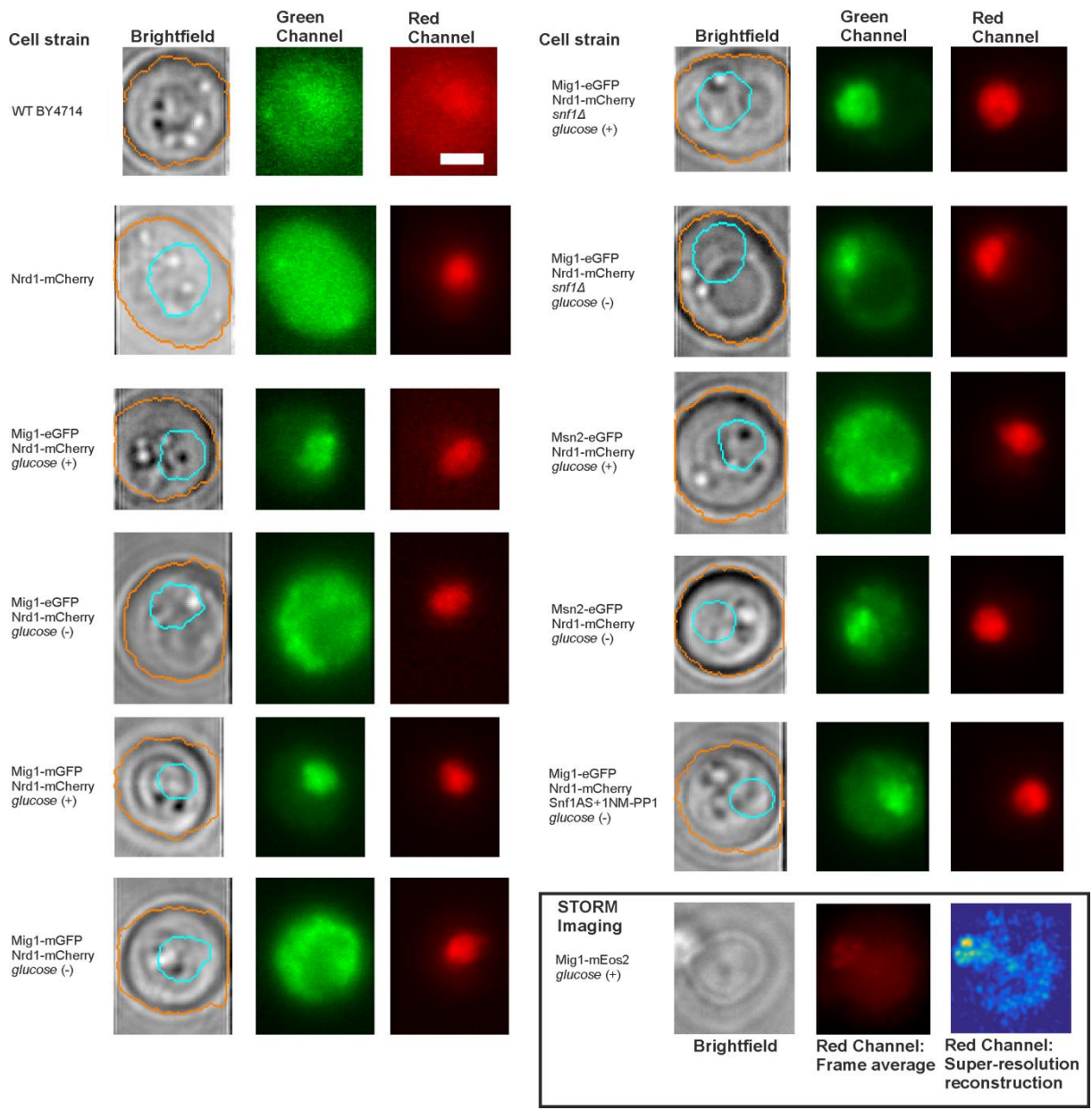

**Fig. S1. Brightfield and fluorescence micrographs of key strains and glucose conditions.**
Representative Slimfield fluorescence images obtained from the strains and different extracellular glucose conditions used in this study. Brightfield non-fluorescence images, segmentation perimeter indicated for cell body (orange) and nucleus (cyan), and fluorescence images are indicated, the latter showing both green and red channels obtained as the frame average from the first five consecutive Slimfield images. Fluorescence images are all normalized by total pixel intensity. For the Mig1-mEos2 strain (inset, bottom right) this shows the brightfield image (left panel), a 300 consecutive frame average from the red channel after photoconversion (middle panel) and super-resolution false color heat map reconstruction, 40nm lateral resolution, >2,000 localizations (right panel).

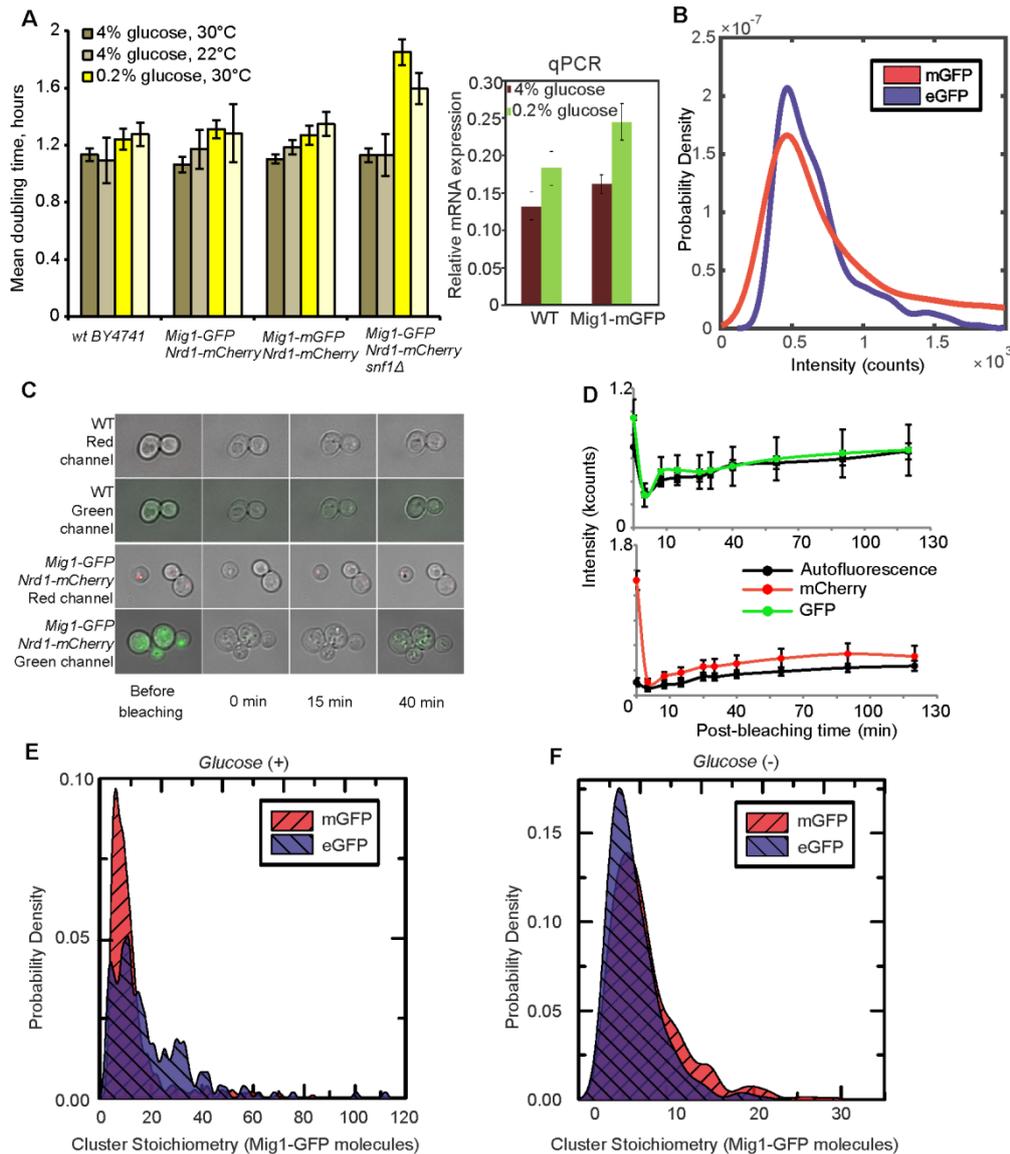

**Fig. S2. Fluorescent reporter strains have similar viability to wild type, with relatively fast maturation of fluorescent protein, and no evidence for GFP-mediated oligomerization.**

(**A**) (left panel) Mean doubling time ± SEM, number of cultures n=6; (right panel) relative expression of *MIG1* to constitutive *ACT1* using qPCR in the wild type and Mig1-mGFP in cells pre-grown in elevated (4%) and depleted (0.2%) glucose, SD error bars, n=3 repeats for each. (**B**) 'Monomeric' mGFP (red) *vs* Standard enhanced eGFP (blue) *in vitro* intensity distributions. GFP/mCherry maturation. (**C**) After continuous illumination images were taken at subsequent time intervals. To prevent appearance of newly synthesized fluorescent proteins, 100 µg/ml cycloheximide was added 1h prior to photobleaching. Upper panels represent autofluorescence appearance in green and red channels in BY4741 wild type cells. (**D**) GFP and mCherry maturation in minimal YNB media with complete amino acid supplement and 4% glucose. The background-corrected total cellular fluorescence intensity for the wild type (autofluorescence) and Mig1-GFP:Nrd1-mCherry strain was quantified at each time point for each cell in ImageJ. Error bars indicate SEM. (**E**) **and** (**F**) *In vivo* Mig1-GFP *vs* Mig1-mGFP stoichiometry distributions compared in *glucose* (+) and *glucose* (-) respectively.

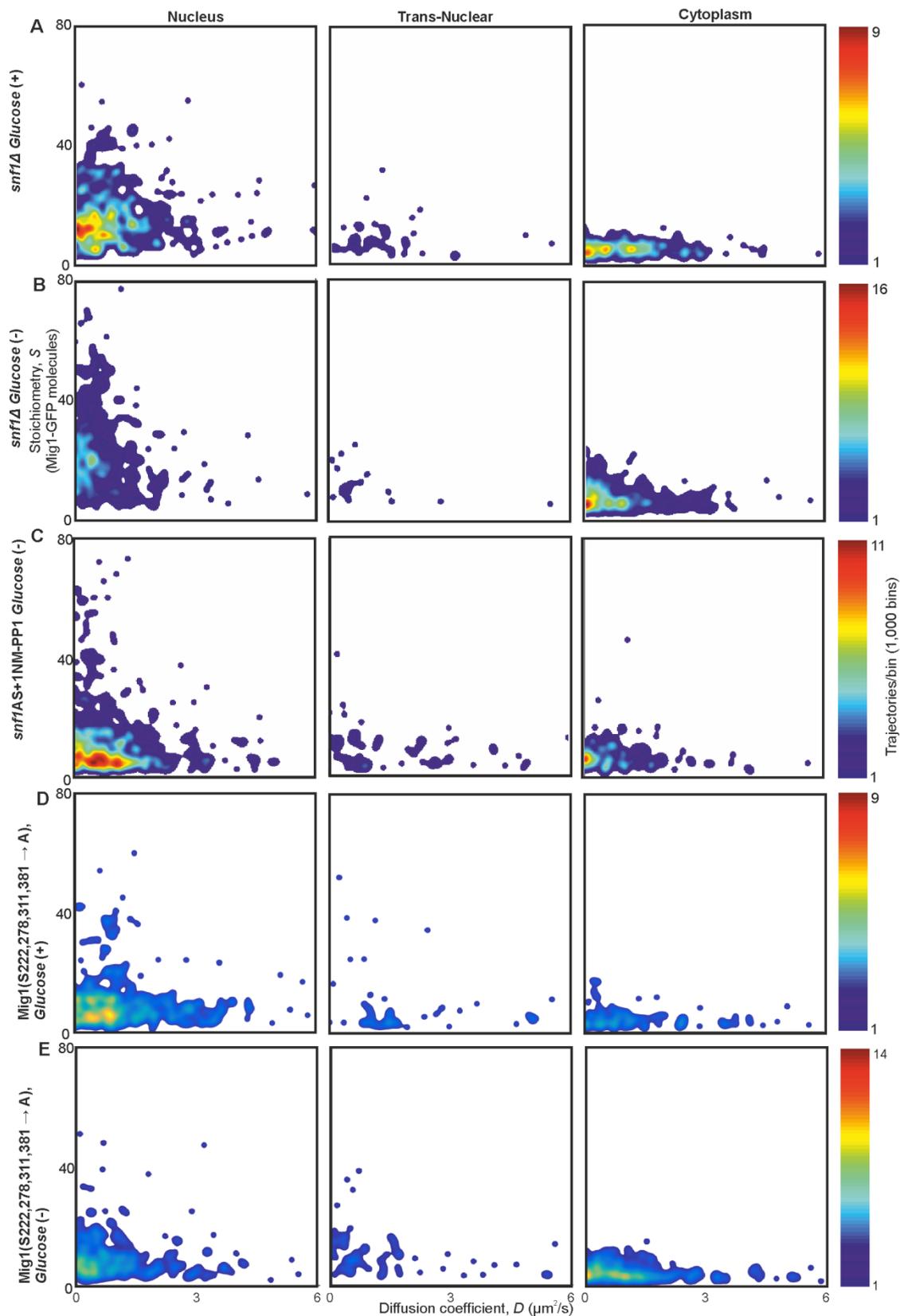

**Fig. S3. Mig1 phosphorylation does not affect clustering but regulates localization.** Heat maps showing dependence of stoichiometry of detected GFP-labeled Mig1 foci with $D$ in (**A**, **B**) *SNF1* deletion strain, (**C**) strain with ATP analog sensitive variant of Snf1 in presence of 1NM-PP1, and (**D**, **E**) strain with four serine phosphorylation sites of Mig1 mutated to alanine.

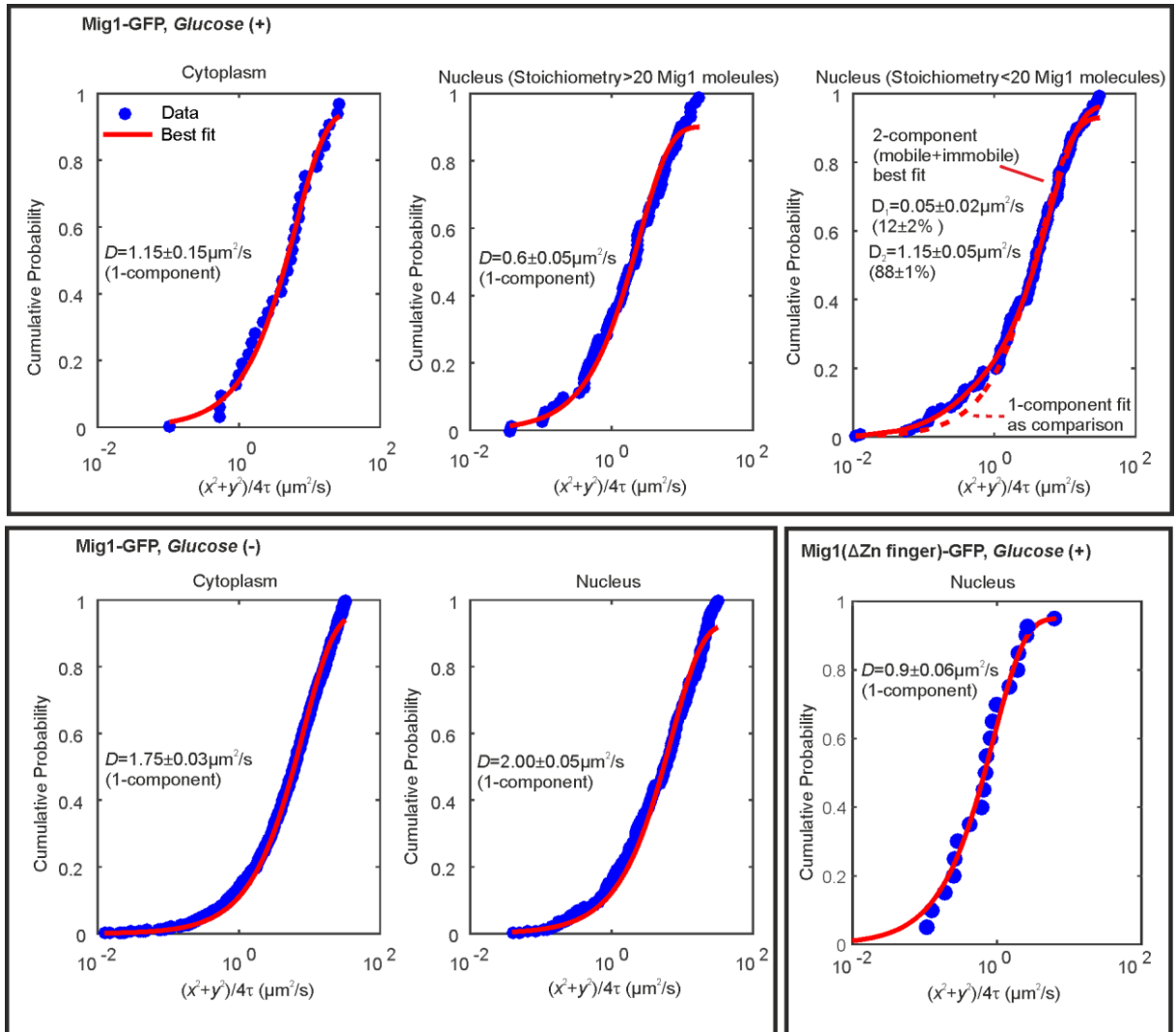

**Fig. S4. Cumulative probability distance analysis reveals a single mobile population in the cytoplasm at glucose (+/-) and in the nucleus and glucose (-).** Cumulative density functions of first displacement in trajectories (blue) with appropriate fits (red). Bottom right panel indicates Mig1 mutant for which the Zn finger domain has been deleted.

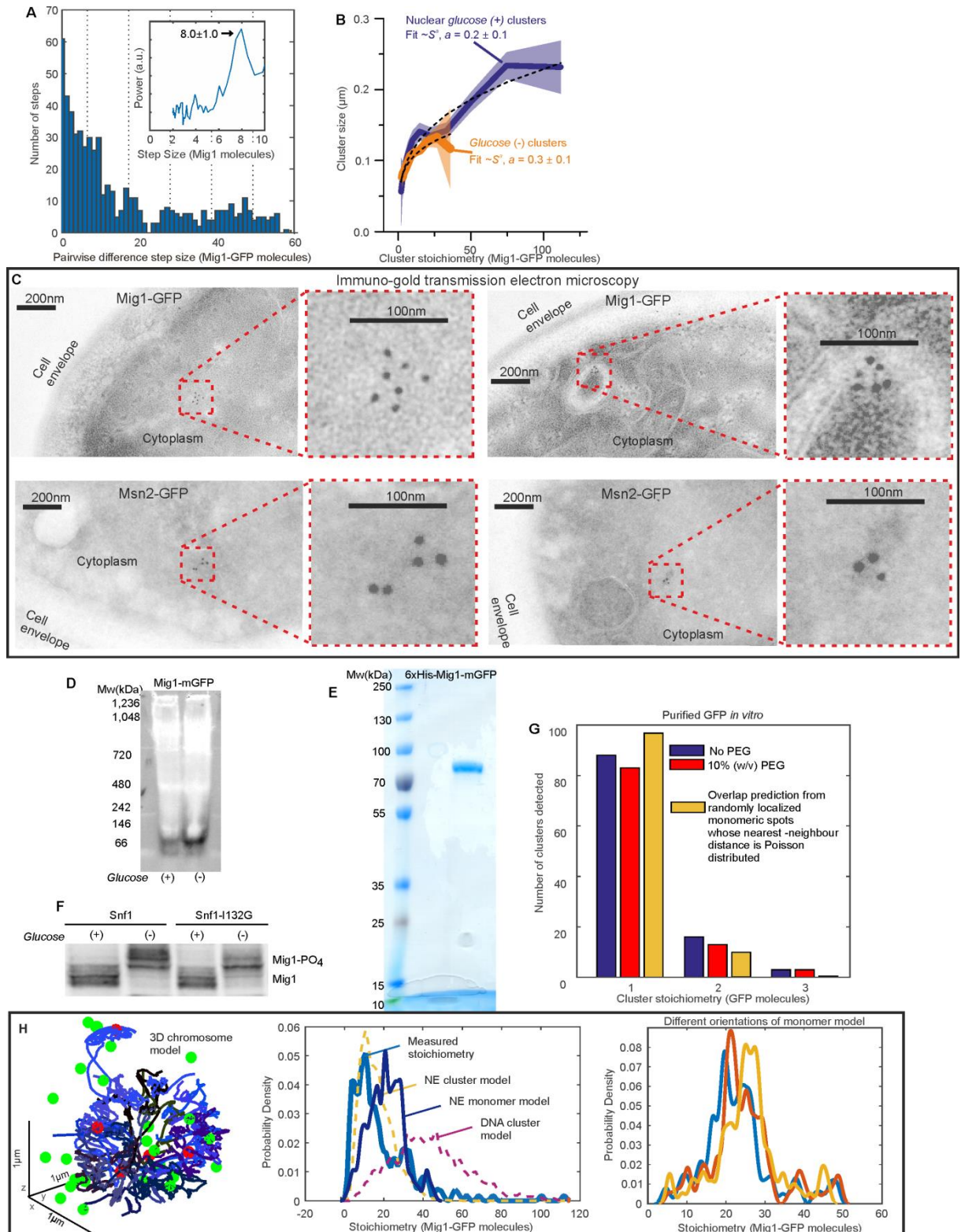

**Fig. S5. Additional Mig1 cluster investigations.** (**A**) Zoom-in on pairwise difference distribution for stoichiometry of GFP-labeled Mig1 foci FRAP experiments, ~8-mer intervals (dashed lines) and power spectrum (inset) shown, mean and Gaussian sigma error (arrow). (**B**) GFP-labeled Mig1 cluster size as a function of stoichiometry with power law fit indicated. (**C**) Immuno-gold transmission electron microscopy for negatively stained 90nm cryosection of (upper panel) two different Mig1-GFP cells and (lower panel) two different Msn2-GFP cells, with zoom in inset. (**D**) Native PAGE of total cell protein extracts obtained

from cells grown in 4% *glucose* (+) and 0.05% *glucose* (-) conditions followed by western blotting and probing with anti-GFP antibodies. **(E)** Coomassie staining of purified Mig1-mGFP fraction indicates a single band that corresponds to the size of a Mig1-GFP monomer (molecular weight 83.4kDa). **(F)** Mig1 phosphorylation status is detected by SDS-PAGE on total cell protein extracts obtained from cells grown in different glucose conditions followed by western blotting. De/phosphorylation of Mig1 in *glucose* (+/-) respectively is not affected by the *SNF1- I132G* mutation. **(G)** Distribution of stoichiometry for mGFP clusters *in vitro* in absence (blue)/presence (red) of PEG and the expected distribution of overlapping mGFP monomers (yellow). **(H) (left panel)** 3C model (blue) with overlaid bound Mig1 clusters to promoter binding sites from bioinformatics (red), and Mig1 clusters near the NE (green); **(middle panel)** predicted stoichiometry distributions for GFP-labeled Mig1 foci in the nucleus at elevated extracellular glucose for a range of different binding models, including: a model which simulates both nuclear envelope (NE) translocating clusters and cluster binding to promoter targets (yellow), a model which simulates both nuclear envelope (NE) translocating monomers and monomer binding to promoter targets and DNA (blue), and a model which simulates just cluster binding to promoter targets but excludes any effects from translocating clusters (purple). These models are optimized to the experimentally determined stoichiometry distribution (cyan); **(right panel)** predicted Mig1 monomer stoichiometry distributions for Mig1 bound to promoter sites in three different orientations ~10º apart.

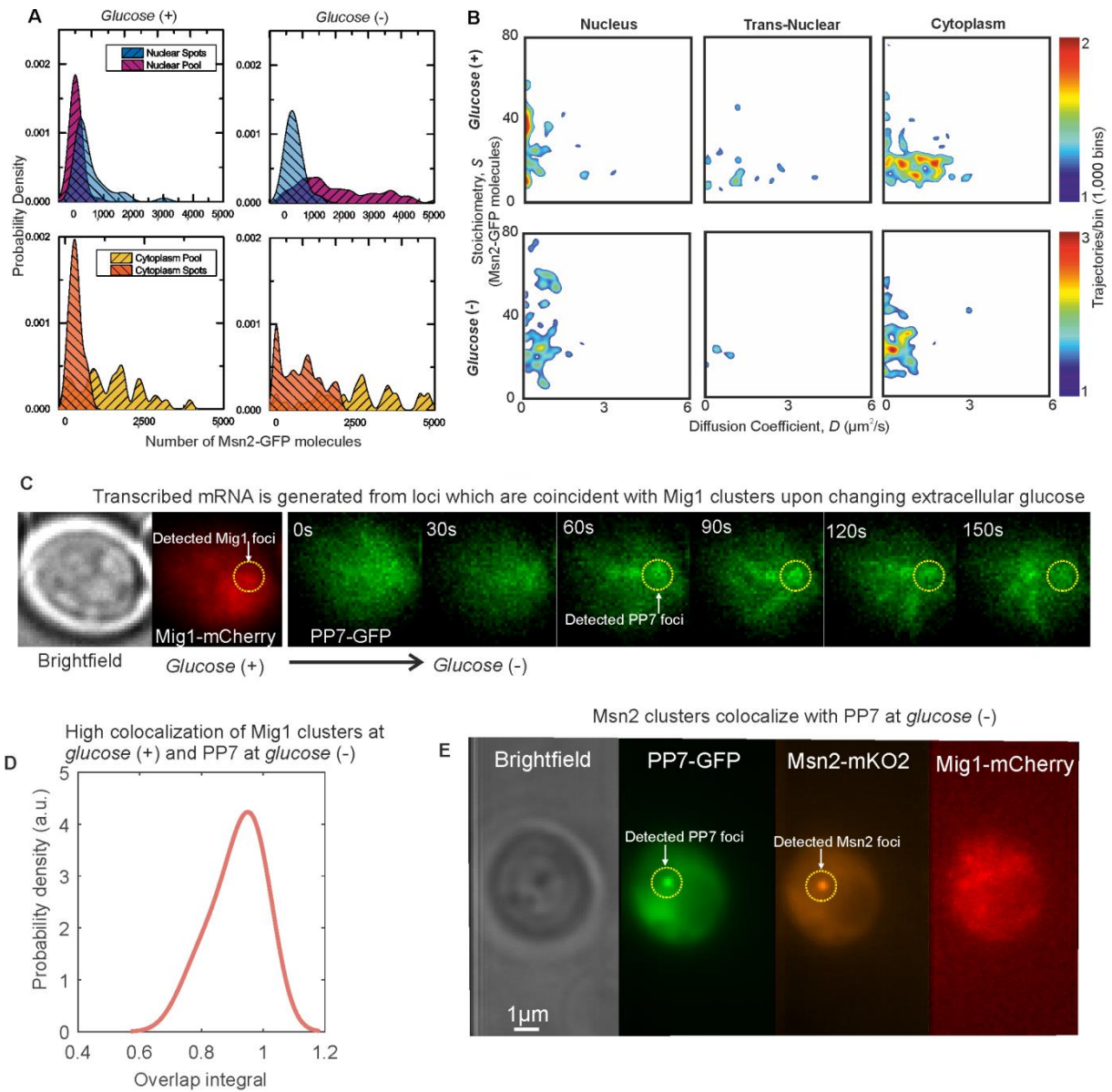

**Fig. S6. Msn2 and Mig1 forms functional clusters colocalized to transcribed mRNA from their target genes (A)** Kernel density estimations for Msn2-GFP in pool and foci for cytoplasm and nucleus at *glucose* (+/-). (**B**) Heat maps showing dependence of stoichiometry and $D$ of detected Msn2-GFP foci, n=30 cells. (C) Slimfield imaging on the same cell in which microfluidics is used to switch from *glucose* (+) to *glucose* (-) indicating the emergence of PP7-GFP foci at *glucose* (-) which are coincident with Mig1-mCherry foci at *glucose* (+). These Mig1 and PP7 foci have a high level of colocalization as seen from **(D)** the distribution of the numerical overlap integral between foci in red and green channels at *glucose* (+) and *glucose* (-) respectively, peaking at ~0.95. **(E)** At *glucose* (-) Msn2-mKO2 foci colocalize with PP7-GFP foci.